\documentstyle[11pt]{article}

\input{tcilatex}
\input epsf
\begin{document}

\font\cmss=cmss10 \font\cmsss=cmss10 at 7pt 
\hfill CERN-TH/99-97

\hfill

\vspace{20pt}

\begin{center}
{\Large {\bf {Quantum Irreversibility in Arbitrary Dimension}}}
\end{center}

\vspace{6pt}

\begin{center}
{\sl Damiano Anselmi}

{\it CERN, Division Th\'eorique, CH-1211, Geneva 23, Switzerland }
\end{center}

\vspace{12pt}

\begin{center}
{\bf Abstract}
\end{center}

\vspace{4pt}{\small Some recent ideas are generalized from four dimensions
to the general dimension $n$. In quantum field theory,
two terms of the trace anomaly in external
gravity, the Euler density ${\rm G}_n$ and $\Box ^{n/2-1}R$, are relevant to
the problem of quantum irreversibility. By adding the divergence of a
gauge-invariant current, ${\rm G}_n$ can be extended to a new notion
of Euler density $\tilde{{\rm G}}_n$, linear in the conformal factor.
We call it {\it pondered} Euler density. This notion
relates the trace-anomaly coefficients $a$ and $a^{\prime }$ of ${\rm G}_n$
and $\Box ^{n/2-1}R$ in a universal way ($a=a^{\prime }$) and gives a
formula expressing the total RG flow of $a$ as the invariant area of the
graph of the beta function between the fixed points. I illustrate these
facts in detail for $n=6$ and check the prediction to the fourth-loop order
in the $\varphi ^{3}$-theory. The formula of quantum irreversibility for
general $n$ even can be extended to $n$ odd by dimensional continuation.
Although the trace anomaly in external gravity is zero in odd dimensions, I
show that the odd-dimensional formula has a predictive content. }

\vskip 9.5truecm \noindent CERN-TH/99-97 -- April, 1999

\vfill\eject 

\section{Introduction}

\setcounter{equation}{0}

The trace anomaly is a useful tool to study non-perturbative properties of
quantum field theory making use of perturbation theory only. A particularly
fruitful arena for this kind of investigation is the conformal window, where
the renormalization-group (RG) flow interpolates between two conformal field
theories; this interpolation is understood in the sense of the resummation
of the perturbative expansion, even when the interacting fixed point is
strongly coupled.

Embedding the theory in external gravity is particularly convenient to study the
RG interpolation between the UV\ and IR values of the so-called {\it central
charges}, i.e. the coefficients $c$, $a$ and $a^{\prime }$ of the
gravitational invariants appearing in the trace anomaly.

The central charges satisfy positivity conditions and the spectra of
anomalous dimensions obey even stronger restrictions \cite{ferra,nach},
which give non-trivial information about the low-energy limit of the theory.
Other inequalities are strictly related to unitarity, such as the
irreversibility of the RG flow, which is 
the statement that $a_{{\rm UV}}\geq a_{{\rm IR}}$.

In \cite{athm} we studied the problem of quantum irreversibility in four
dimensions. A\ non-perturbative formula expressing
the total $a$-flow as the invariant area of the graph of the beta function
between the fixed point comes out naturally. This formula agrees with
perturbation theory up to the fourth-loop order included. Moreover, a 
natural positivity property of the induced action for the conformal factor
implies the $a$-theorem to all orders, i.e. the inequalities $a_{{\rm UV}%
}\geq a_{{\rm IR}}\geq 0$ and the formula for $\Delta a=a_{{\rm UV}}-a_{{\rm %
IR}}$.

Precisely, it was explained  in \cite{athm} that quantum irreversibility is
measured by the invariant area of the graph of the beta function between the
fixed points. In particular, there exists a one-form 
\begin{equation}
\omega =-{\rm d}\lambda ~\beta (\lambda )~f(\lambda ),  \label{omega}
\end{equation}
where $f(\lambda )\geq 0$ is a metric in the space of coupling constants,
defined by 
\[
\langle \Theta (x)~\Theta (0)\rangle =\frac{1}{15\pi ^{4}}\frac{\beta
^{2}(t)f(t)}{|x|^{8}}
\]
in four dimensions ($t=\ln |x|\mu $). $\Theta$ is the trace of the stress-tensor.
The ideas of \cite{athm} can be
summarized by the formula 
\begin{equation}
\Delta a=\Delta a^{\prime }=\int_{{\rm UV}}^{{\rm IR}}\omega \geq 0,
\label{formula}
\end{equation}
saying that the total RG\ flow of $a$ is the integral of the form $\omega $
between the fixed points. This statement was checked to the fourth-loop order
in the most general renormalizable four-dimensional theory.

It is therefore interesting to understand whether the new approach extends
directly to arbitrary dimension, or further obstructions, similar to the conceptual gap between
irreversibility in two- and four-dimensions, make this task less straightforward.

However, the motivation for exploring higher-dimensional field theories
is not purely academic. The final aim is to extend
the study of quantum irreversibility, well-established in two and four
dimensions, to three-dimensional quantum field theory and phenomena relevant
to condensed matter physics, for instance superconductivity.

The trace anomaly is an intrinsic effect of renormalization, in particular a
non-vanishing beta function \cite{adler,nielsen,collins}, and exists
in every dimension. For example, $\Theta $ is ${\frac{\beta (\lambda )}{6!}}%
\varphi ^{6}$ in the $\varphi ^{6}$-theory in three dimensions. The anomaly
begins at two loops instead of one (because odd-loop Feynman graphs do not
diverge in odd dimensions), which is the reason why it is not visible as an
ordinary one-loop anomaly.

The trace anomaly operator equation in flat space is not sufficient in
itself to study the phenomenon of quantum irreversibility. One needs the
trace anomaly in the presence of an external background (gravity in even
dimensions) that supports certain invariants (such as the Euler density) and
therefore appropriate anomalies, the central charges, in particular the
quantities $a$ and $a^{\prime }$. There is no such background in odd
dimensions. Therefore, we do not see any other way of extending the ideas to
odd dimensions than continuing those formulas that hold in even dimensions.
From the formal point of view, this does not present problems. The
coefficients to be continued to odd dimensions are indeed very simple
functions of the dimension $n$. I will discuss a situation in which the
odd-dimensional formula can in principle be checked.

This is, in summary, the reason why it is interesting to treat the problem
of generic $n$. I start by presenting some aspects of the ideas of \cite
{athm} from the point of view of differential geometry, to clarify how the
generalization to arbitrary $n$ can be achieved. I stress, however, that the
major source of inspiration for these ideas is physics. 
Many identities could be found only with a major effort, if they
were not inspired by the physical ideas behind them.

For definiteness, in the first part of the paper I focus on $n=6,$ or $%
n=6-\varepsilon $. Several identities hold for arbitrary $n$, a fact that is
quite important to reduce the number of independent terms in the regularized
theory. A special section is devoted to the generalization of the 
irreversibility formula to arbitrary dimension.

The Gauss--Bonnet integrand, or Euler density, 
\begin{equation}
{\rm G}_{6}=-\varepsilon ^{\mu \nu \rho \sigma \tau \xi }\varepsilon _{\alpha
\beta \gamma \delta \epsilon \zeta }R_{\mu \nu }^{\alpha \beta }R_{\rho
\sigma }^{\gamma \delta }R_{\tau \xi }^{\epsilon \zeta },  \label{G6}
\end{equation}
is a non-trivial total derivative, i.e. the total derivative of a
non-gauge-covariant current (the Chern--Simons form). It is clear, then,
that ${\rm G}_{6}$ is defined up to trivial total derivatives, i.e. up to
the divergence of a gauge-covariant current. The topological numbers
calculated with any modified Gauss--Bonnet integrand of the form $\tilde{%
{\rm G}}_{6}={\rm G}_{6}+\nabla _{\alpha }J^{\alpha }$ are exactly the same
as those computed with the ordinary one. The modified integrand, however,
can be chosen so as to have an additional remarkable property, namely to be
linear in the conformal factor. The anomaly equation and the induced action
for the conformal factor simplify enormously. The modified integrand will be
called {\it pondered} Euler density and is meaningful in every even dimension.

This phenomenon is already apparent in four dimensions, where the 
pondered Euler
density\footnotemark 
\footnotetext{
In the notation of the present paper ${\rm G}_{4}$ is 4 times the
Gauss--Bonnet integrand of ref. \cite{athm}. See (\ref{definition}).} is 
\begin{equation}
\tilde{{\rm G}}_{4}={\rm G}_{4}-{\frac{8}{3}}\Box R={\rm G}_{4}+\nabla
_{\alpha }J_{4}^{\alpha },~~~~~~~J_{4}^{\alpha }=-{\frac{8}{3}}\nabla
_{\alpha }R.  \label{G4}
\end{equation}
According to ref. \cite{athm}
the conceptual gap between quantum irreversibility in
two \cite{zamolo} and four dimensions is filled by the pondered
extension of the Euler density. {\it This} is the
true Euler density that should appear in the trace anomaly operator
equation, thereby removing the ambiguities associated
with the trivial total derivative terms appearing in the trace anomaly. 
The higher-dimensional extension of these ideas requires 
no other conceptual innovation.

In higher dimensions the possibility of fine-tuning the coefficients of the
numerous additional terms to construct a pondered Euler density is far more
upsetting than in four dimensions, in view of the complexity of the
expressions.

Inspired by these ideas, we have found that in six dimensions the
combination 
\begin{equation}
\sqrt{g}\,\tilde{{\rm G}}_{6}\equiv \sqrt{g}\left[ {\rm G}_{6}+\nabla
_{\alpha }J_{6}^{\alpha }\right] =48~\Box ^{3}\phi   \label{six}
\end{equation}
is indeed linear in the conformal factor $\phi $ for conformally flat
metrics $g_{\mu \nu }={\rm e}^{2\phi }\delta _{\mu \nu }$, if 
\begin{eqnarray*}
J_{6}^{\alpha } &=&-\left( \frac{408}{5}-20\zeta \right) R_{\mu }^{\nu
}\nabla _{\nu }R^{\mu \alpha }-\left( \frac{36}{25}+2\zeta \right) R^{\alpha
\mu }\nabla _{\mu }R \\
&&+~\zeta \nabla ^{\alpha }R^{2}+\left( \frac{144}{5}-10\zeta \right) \nabla
^{\alpha }(R_{\mu \nu }R^{\mu \nu })-\frac{24}{5}\nabla ^{\alpha }\Box R.
\end{eqnarray*}
We do not need to use the Riemann tensor $R_{\mu\nu\rho\sigma}$. 
We re-express it as a combination of the Ricci 
tensor $R_{\mu\nu}$ and the scalar curvature $R$, plus the Weyl tensor $W_{\mu\nu\rho\sigma}$. 
Most of our work in this paper
focuses on the trace $\Theta$ of the stress-tensor, coupled to the conformal factor $\phi$.
It is understood that our formulas are written ``up to terms
proportional to the Weyl tensor''.

Our requirement (\ref{six}) does not seem to fix $J_6^{\alpha }$ uniquely, but leaves a
free parameter $\zeta $. We do not need $\zeta $, since $\zeta $ does not
affect the term $\Box ^{2}R$, but this parameter is the sign of a redundancy
in our list of terms. Exploring this issue better, we find that the
combination 
\begin{equation}
\nabla _{\alpha }\left[ \nabla ^{\alpha }R^{2}-2(n-1)\nabla ^{\alpha
}(R_{\mu \nu }R^{\mu \nu })-2R^{\alpha \mu }\nabla _{\mu }R+4(n-1)R_{\mu
}^{\nu }\nabla _{\nu }R^{\mu \alpha }\right]   \label{iden}
\end{equation}
is identically zero on conformally-flat metrics in arbitrary dimension $n$.
The fact that this is true for arbitrary $n$ and not just for $n=6$ allows
us to consistently reduce the number of terms, in a regularized theory as
well. We choose $\zeta ={\frac{102}{25}}$, so that 
\begin{equation}
J_{6}^{\alpha }=-\frac{48}{5}R^{\alpha \mu }\nabla _{\mu }R+{\frac{102}{25}}%
\nabla ^{\alpha }R^{2}-12\nabla ^{\alpha }(R_{\mu \nu }R^{\mu \nu })-\frac{24%
}{5}\nabla ^{\alpha }\Box R.  \label{jeia}
\end{equation}
Writing $\sqrt{g}{\rm G}_{6}=\partial _{\alpha }C^{\alpha },$ where $%
C^{\alpha }$ is the Chern--Simons form, the {\it pondered} Chern--Simons
form reads $\tilde{C}^{\alpha }=C^{\alpha }+\sqrt{g}J^{\alpha }$ and $\sqrt{g%
}{\rm \tilde{G}}_{6}=\partial _{\alpha }\tilde{C}^{\alpha }.$

In ref. \cite{armeni} Karakhayan {\it et al.} have explicitly
written a conformal-invariant operator of the form $\Box ^{3}+$curvature
terms in six dimensions. A similar operator (of the form $\Box ^{2}+$%
curvature terms) exists in four dimensions \cite{riegert}. It is actually
the variation of the pondered Euler density with respect to the conformal
factor, and it can be read in six dimensions by 
varying $\tilde {\rm G}_6$ under a
general rescaling $g_{\mu \nu }\rightarrow {\rm e}^{2\phi }g_{\mu \nu }$.
$\tilde {\rm G}_6$ can also be derived from the results of ref. 
\cite{armeni}, with a certain amount of work.
The pondered Euler density is the basic ingredient to write down the
higher-dimensional analogue of the Riegert action \cite{riegert}; it is
unique up to terms proportional to the Weyl tensor. In section 5 the
expression of ${\rm \tilde{G}}_{8}$ is reported and the construction
for arbitrary even $n$ is presented.

The notion of pondered Euler density relates in a universal way the
trace-anomaly coefficients of ${\rm G}_{6}$ and $\Box ^{2}R$, which we can
call $a$ and $a^{\prime }$, respectively. It allows us to generalize the
formula of \cite{athm}, measuring the effect of quantum irreversibility in
terms of the area of the graph of the beta function between the fixed points
(section 2). 

The rest of the paper is organized as follows.
After a detailed classification of the curvature terms
occurring in the bare lagrangian and in the trace anomaly (section 3), we
check this prediction to the fourth-loop order in perturbation theory for
the $\varphi ^{3}$-theory in six dimensions (section 4); this is a good
opportunity to show that the RG equations produce exactly the pondered Euler
density, eliminating step by step the $\phi $-interaction terms in the trace
anomaly. The $a$-theorem can indeed be seen as a non-renormalization theorem
for the $\Theta $ many-point functions. The general formula is extended to
arbitrary $n$ in section 5, where the continuation to odd dimensions is
discussed, together with a suggestion for a check.

Before beginning the technical study, we make a comment of a general
character. Quantum irreversibility is intrinsically related to the beta
function, as we have stressed. The beta function plays a major role
in the RG interpolation problem. Now,
perturbation theory cannot be applied too
naively in this context, otherwise
fake ambiguities and sometimes wrong conclusions do arise. There is a
well-known scheme change, proposed by 't Hooft \cite{scheme}, in which the
beta function is just two-loop. If this were consistent in the
RG-interpolation problem, we could conclude 
immediately that the $\varphi ^{4}$-theory, say,
has a UV interacting fixed point, that pure Yang--Mills theory has no IR
fixed point, etc. However, the scheme change under consideration is allowed 
only order by order in perturbation theory.

The point is that an apparently harmless power series in $\alpha $ can resum
to a function proportional to the inverse of the beta function. An example
is the function $\alpha /\beta $ in QCD, where $\beta ={\rm d}\ln \alpha /%
{\rm d}\ln \mu $. A scheme change generated by such a function is admissible
in the naive sense, not in our problem, since it spoils a proper
interpolation between the fixed points. The moral of the story is that one
has to keep track of the orders of $\beta $ {\it and} $\alpha $, not only of the
powers of $\alpha $ \footnote{%
These ideas have been discussed and applied extensively in \cite{ccfis,noi},
to which we refer the readers for further details.}.

The appearance of orders of $\beta $ can often be rigorously detected and a
good approach to the RG interpolation problem relies on this fact. An
illustration is provided precisely by the trace anomaly, whose existence was
first realized by Coleman and Jackiw in ref. \cite{CJ}. It was later shown 
\cite{adler} that the general form of the trace anomaly reads $\Theta =\beta 
{\rm O}$, ${\rm O}$ being a certain operator ($F^{2}/4$ in the case of QED\
and Yang--Mills theory). While both $\beta $ and ${\rm O}$ depend on the
scheme, the invariant content of this equation is precisely that $\Theta $
is of order $\beta $ around the fixed points and that a scheme change that
spoils this property is not admissible. This is in agreement with the
restored conformal invariance at the fixed points.

Sometimes, axial anomalies are proportional to the beta function also. In
particular, this applies, in supersymmetric theories, to the divergence of
the $R$-current, which is the supersymmetric partner of $\Theta $. There is
no contradiction with the Adler--Bardeen theorem, since currents differring
by ${\cal O}(\epsilon )$-terms (we assume that we are working in dimensional
regularization) can well satisfy different anomaly equations. In one case
the divergence has the form predicted by the Adler--Bardeen theorem, in
another case it is proportional to the beta function \cite{zanon}. In
agreement with our observations, the scheme change that interchanges
the two anomaly equations is singular (see for example ref. \cite{rattazzi}%
). The important thing for the applications \cite{noi} is to identify which
current is the correct partner of $\Theta $ (this problem has been known for
some time as the {\sl anomaly puzzle}).

When embedding the theory in external gravity, the careful analysis of \cite
{hathrell2,hathrell}, which we are going to use extensively in this paper,
identifies other places where orders of $\beta $ appear. Most scheme
dependences in the central charges are proportional to the beta function 
\cite{ccfis} and therefore do not affect the critical values and the total
RG\ flows. These techniques allow us to show that the renormalization-group
equations select precisely the combination of gravitational invariants that
we have called pondered Euler density.

In summary, quantum field theory defines a natural fibre bundle, 
that we call {\it scheme bundle}. The base
manifold is the space of physical correlators and the fibre is the space of
allowed scheme choices, with the regularity restriction outlined above. 
A projection onto the base manifold is defined and assures scheme independence
of the physical correlators.
The scheme bundle is equipped with a metric $f$ and a fundamental one-form $\omega $,
see (\ref{omega}). In integrals expressing total RG flows, such as (\ref
{formula}), scheme independence is reparametrization invariance. The bundle
admits a ``proper'' section (and so a proper beta function and a proper
coupling constant), defined as the scheme choice for which the metric $f$ is
constant throughout the RG flow (see \cite{athm} for the detailed
construction).

This geometric approach to quantum field theory is complementary to a
similar notion introduced in \cite{reietto} for the algorithm of subtraction
of divergences in the most general gauge field theory. In that case a fibre
bundle was defined, whose fibre was the space of fields and antifields and
whose base manifold was, again, the space of physical parameters and correlators.
The removal of divergences was proved to be the direct product of a diffeomorphism in the base
manifold and a canonical transformation on the fibre. 
There is a natural structure of connections and curvatures that
helps the generalization of the known theorems of renormalization theory.

\section{Idea and prediction}

\setcounter{equation}{0}

At criticality the trace anomaly reads 
\begin{equation}
\Theta =a_{*}\ {\rm G}_{6}+{\rm %
conformal~invariants}+{\rm trivial\,divergences}.  \label{critica}
\end{equation}
For a free theory with $%
N_{s}$ real scalar fields, $N_{f}$ Dirac fermions and $N_{v}$ 2-forms,
we have 
\begin{equation}
a_{\rm free}=(N_{s}+f_{6}N_{f}+v_{6}N_{v})\, {\rm const.}  \label{free}
\end{equation}
Results for $a_{\rm free}$, in particular the overall constant 
and the relative factor $f_6$, can be found in ref. \cite{ichinose}.

The idea of \cite{athm} is that the ``trivial'' divergences are fixed
uniquely by the property (\ref{six}) and in the end are not so trivial as is
commonly assumed, since they produce a precise formula for the effect of
quantum irreversibility. From the point of view of explicit calculations,
these terms contain a certain type of ambiguity, which makes them
ill-defined by arbitrary additive finite bare parameters. 
Nevertheless, the ambiguities can be consistently removed in
a universal way, writing
\begin{equation}
\Theta =a_{*}\ \tilde{{\rm G}}_{6}+{\rm %
conformal\,invariants}.  \label{critica2}
\end{equation}

The two-point function of $\Theta $ is proportional to the number $a_{*}$ in
the conformal limit $\beta =0$. We get 
\begin{equation}
\langle \Theta (x)\ \Theta (y)\rangle =-\left. \frac{\delta ^{2}S_{{\rm E}%
}[\phi ]}{\delta \phi (x)\delta \phi (y)}\right| =-48\,
a_{*}\ \Box ^{3}\delta (x-y).  \label{aprimo}
\end{equation}
Here $S_{{\rm eff}}[\phi ]$ denotes the induced effective action for the
conformal factor. The notation $\left. Q\right| $ means that the quantity $Q$ is evaluated in
flat space. Formula (\ref{critica}) gives, in the case of a conformally-flat
metric, 
\[
\Theta =48\, a_{*}\Box ^{3}\phi ,
\]
and the induced action is 
\[
S_{{\rm E}}=-24\,a_{*}\int {\rm d}%
^{6}x\,(\Box \partial _{\mu }\phi )^{2}.
\]

With the identification $a^{\prime }=a$, i.e. with the modified
Gauss--Bonnet integrand, the $\phi $-action is free at criticality, which means that the
three- and many-point functions of $\Theta $ are zero. In
four dimensions the two-point function $\langle \Theta (x)~\Theta (0)\rangle 
$ equals $-\frac{1}{90(4\pi) ^{2}}a\Box ^{2}\delta (x)$. In two dimensions
it is $\langle \Theta (x)~\Theta (0)\rangle =-{\frac{\pi }{3}}c\Box \delta
(x).$

Using formula (\ref{aprimo}) we see that the quantity $a_{*}$ can be
expressed at criticality by the integral 
\begin{equation}
a_{*}=-\frac{1}{2^{13}\cdot 3^3\cdot 5}
\int {\rm d}^{6}x\ |x-y|^{6}\langle \Theta (x)\
\Theta (y)\rangle .  \label{ugh}
\end{equation}

We now consider the off-critical theory. We can define a function $a^{\prime
}(r)$ of the intermediate energy scale $1/r$ by restricting the integration
over a four-sphere $S(r,y)$ of radius $r$ and centred at the point $y$,
precisely 
\begin{equation}
a(r_{2})-a(r_{1})=-\frac{1}{2^{13}\cdot 3^3\cdot 5}
\int_{S(r_{1},y)}^{S(r_{2},y)}{\rm d}%
^{6}x\ |x-y|^{6}\langle \Theta (x)\ \Theta (y)\rangle .  \label{ar}
\end{equation}
For a critical theory we have $a_{{\rm UV}}=a_{{\rm IR}}=a(r)=a_{*}.$
Off-criticality, the running of $a^{\prime }(r)$ is due to the internal term 
(e.g. $\frac{\beta }{3!}\varphi ^{3}$ in the $\varphi^3$-theory) appearing in the expression of
$\Theta $. There is a non-local contribution to the correlator $%
\langle \Theta (x)~\Theta (y)\rangle $, 
\begin{equation}
\langle \Theta (x)\ \Theta (y)\rangle ={\frac{2^{13}\cdot 3^3\cdot 5}
{\pi ^{3}}}\,\,\frac{\beta
^{2}[\lambda (t)]f[\lambda (t)]}{|x-y|^{12}}~,~~~~~~~~~~{\rm for}~~x\neq y,
\label{tetateta}
\end{equation}
so that the total flow of the quantity $a$ is non-negative and equal to the
invariant area of the beta function: 
\begin{equation}
a_{{\rm UV}}-a_{{\rm IR}}=\int_{-\infty }^{+\infty }{\rm d}t\ \beta
^{2}(t)f(t)=-\int_{\lambda _{{\rm UV}}}^{\lambda _{{\rm IR}}}{\rm d}\lambda
\ \beta (\lambda )f(\lambda )\geq 0.  \label{resu}
\end{equation}

This prediction can be checked to the fourth-loop order in perturbation
theory. We will consider the $\varphi ^{3}$-theory in six dimensions. This
theory is not a meaningful physical theory, but it is the only
renormalizable theory in dimension greater than four. It was observed in 
\cite{athm} that formula (\ref{resu}) is expected to work even in the
absence of an interacting fixed point and can be consistently checked order
by order in perturbation theory whenever the perturbative expansion makes
sense. The price is that we cannot demand that the inequality $a_{{\rm UV}}\geq a_{%
{\rm IR}}$, which is the natural generalization of Cardy's conjecture \cite
{cardy} to six dimensions, be satisfied.

In our notation the lagrangian is 
\begin{equation}
\int {\cal L}=\int {\rm d}^{6}x\left[ {\frac{1}{2}}(\partial _{\mu }\varphi
^{i})^{2}+{\lambda }\varphi _{1}\varphi _{2}\varphi _{3}\right] ,
\label{act}
\end{equation}
where $i=1,2,3$. We choose this form so as to avoid counterterms linear in $%
\varphi $. We sometimes compactify the notation by writing ${\cal L=}{\frac{1%
}{2}}(\partial _{\mu }\varphi )^{2}+\frac{{\lambda }}{3!}\varphi ^{3}$, but
it should be kept in mind that what we mean is the theory (\ref{act}).

We follow the procedure of \cite{hathrell2}, which we generalize to six
dimensions. It was explained in \cite{athm} that we do not need, to the
order we are interested in, to worry about 
the non-minimal term $R\varphi ^{2}$
in the extension of (\ref{act}) to curved space. The two-loop beta function
is \cite{twobeta}

\[
\beta (\lambda )=-{\frac{1}{2}}{\frac{\lambda ^{3}}{(4\pi )^{3}}}-{\frac{41}{%
36}}{\frac{\lambda ^{5}}{(4\pi )^{6}}}+{\cal O}(\lambda ^{7})\equiv \beta
_{1}\lambda ^{3}+\beta _{2}\lambda ^{5}+{\cal O}(\lambda ^{7}). 
\]
The theory is formally asymptotically free. We have $\Theta =-{\beta
(\lambda )}\varphi _{1}\varphi _{2}\varphi _{3}$ and therefore 
\[
f_{*}={\frac{\pi ^{3}}{2^{13}\cdot 3^3\cdot 5}}
|x|^{12}\left( \langle \varphi _{1}(x)\,\varphi
_{1}(0)\rangle \right) ^{3}={\frac{1}{2^{7}\cdot 3^3\cdot 5\,(4\pi )^{6}}}, 
\]
since the $\varphi $-propagator is $\langle \varphi (x)\,\varphi (0)\rangle
=-1/\Box =1/(4\pi ^{3})\,1/|x|^{4}$.

As was explained in ref. \cite{athm}, our formula predicts the $a$-flow even
in the absence of a rigorous fixed point. However, in order to achieve this
goal, we have to pretend that a fixed point does indeed exist. This can be
done as follows. We write $\beta (\lambda )=\beta _{2}\lambda ^{3}(\lambda
^{2}-\lambda _{{\rm IR}}^{2})+{\cal O}(\lambda ^{7})$ and pretend that $%
\lambda _{{\rm IR}}^{2}=-\beta _{1}/\beta _{2}$ is a small parameter. In
other words, we consider $\beta _{1}$ as an independent, small parameter,
and replace it with $-\beta _{2}\lambda _{{\rm IR}}^{2}$ everywhere. This
procedure is unambiguous as long as we can keep track of the orders of $%
\beta $, not only the orders of $\alpha $, as explained in the introduction.
We will see that there is enough information to carry this over to the end.
Of course, there would be no such nuisance in the presence of a true zero of
the beta function.

We therefore have, to the lowest order: 
\[
a_{{\rm UV}}-a(\lambda _{{\rm IR}})=-\int_{0}^{\lambda _{{\rm IR}}}d\lambda
^{\prime }\,{f}_{*}\beta _{2}\lambda ^{\prime ~3}(\lambda ^{^{\prime
}2}-\lambda _{{\rm IR}}^{2})={\frac{1}{12}}\beta _{2}f_{*}\lambda _{{\rm IR}%
}^{6},
\]
and the fourth-loop-order prediction 
\[
a(\lambda )=a_{\rm free}-{\frac{1}{12}}\beta _{2}f_{*}\lambda ^{6}{+O(}\lambda ^{8})=a_{\rm free}+{%
\frac{41}{7464960}\frac{\lambda ^{6}}{(4\pi )^{12}}+O(}\lambda ^{8}),
\]
or 
\begin{equation}
\Theta =\left( a_{\rm free}+\frac{41}{7464960}%
\frac{\lambda ^{6}}{(4\pi )^{12}}\right) {\rm G}_{6}+{\rm rest}.
\label{prediction}
\end{equation}

Note that the check of $\Delta a$ is independent of the value $a_{\rm free}$. 
This value is needed to normalize 
$a$ conventionally, but Hathrell's techniques, which we extend here to six
dimensions, give the flow $\Delta a$ directly and are actually unable of
predicting the value $a_{{\rm free}}$. We stress again that the first
relevant radiative correction is fourth-loop in our case, while it is
generically third-loop in four dimensions.

\vskip .5truecm 
\let\picnaturalsize=N

\ifx\nopictures Y\else{\ifx\epsfloaded Y\else\fi
\global\let\epsfloaded=Y \centerline{\ifx\picnaturalsize N\epsfxsize
4.0in\fi \epsfbox{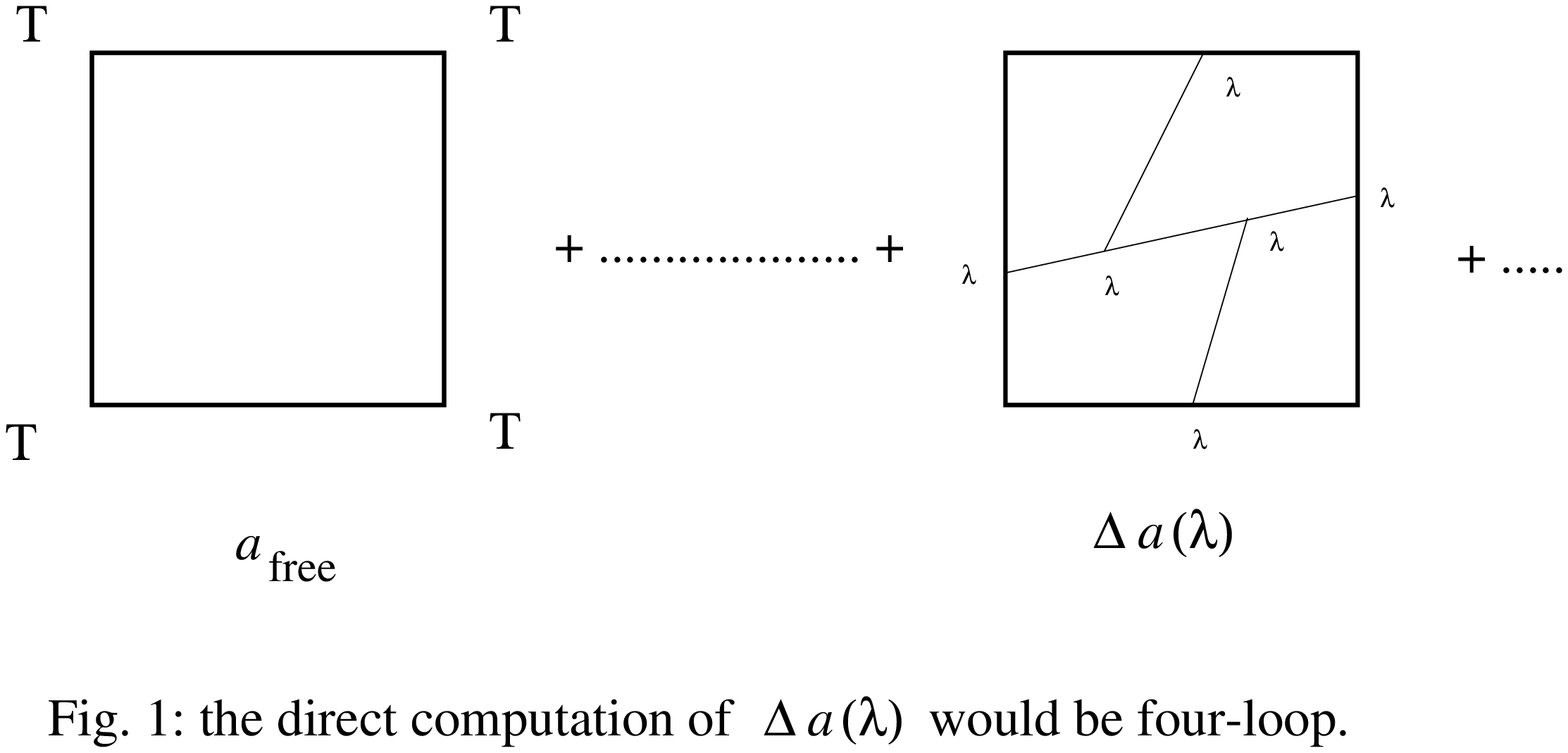}}}\fi

\vskip .5truecm

\section{Classification of the curvature terms}

\setcounter{equation}{0}

In this section we discuss the general structure of the trace anomaly and
the bare lagrangian. A detailed introduction to the subject can be found in
ref. \cite{bonora}.

The conformal-invariant terms are four. Two terms are cubic in the Weyl
tensor, 
\[
A_{1}=W_{\mu \nu \rho \sigma }W^{\mu \nu \alpha \beta }W_{\alpha \beta
}^{\rho \sigma },~~~~~~~~~~~~A_{2}=W_{\mu \nu \rho \sigma }W^{\mu \alpha
\rho \beta }{{{W_{\alpha }}^{\nu }}_{\beta }}^{\sigma }.
\]
A third term has the form 
\begin{equation}
A_{3}=W_{\mu \nu \rho \sigma }\Box W^{\mu \nu \rho \sigma }+\cdots 
\label{WBW}
\end{equation}
and is directly related to the central extension $c$ of the OPE algebra (see 
\cite{ccfis} for details): 
\[
\langle T_{\mu \nu }(x)\,T_{\rho \sigma }(0)\rangle =c\,{\prod }_{\mu \nu
,\rho \sigma }^{(2)}\Box \left( {\frac{1}{|x|^{6}}}\right) ,
\]
$T_{\mu \nu }$ denoting the stress-tensor and ${\prod }_{\mu \nu ,\rho
\sigma }^{(2)}$ being the spin-2 projector (see for example \cite{high}). To
see this, we use the relations 
\[
\mu {\frac{\partial }{\partial \mu }}\rightarrow \int \Theta ,~~~~~~~~~~~\mu 
{\frac{\partial }{\partial \mu }}{\frac{1}{|x|^{6}}}=\pi ^{3}\delta (x),
\]
so that 
\[
\left. {\frac{\delta ^{2}}{\delta g_{\mu \nu }(x)\delta g_{\mu \nu }(0)}}%
\int \Theta \,\,\right| =\left\langle \int \Theta \ T_{\mu \nu }(x)\,T_{\rho
\sigma }(0)\right\rangle \rightarrow \pi ^{3}c\,{\prod }_{\mu \nu
,\rho \sigma }^{(2)}\Box \delta (x)
\]
and therefore $\Theta \sim cA_{3}+{\rm rest}$. The central-extension term
reads $W_{\mu \nu \rho \sigma }\Box ^{n/2-2}W^{\mu \nu \rho \sigma }+\cdots $
in generic even dimension $n> 2$.

The complete form of $A_{3}$ that can be found in the literature (for
example, in ref. \cite{armeni}, formula (2.18)) is not explicitly
proportional to the Weyl tensor. This is because a spurious contribution
proportional to ${\rm G}_{6}$ is included. It is easy to see, however, that
subtracting this term away, $A_{3}$ can be re-expressed in a form that is
proportional to the Weyl tensor (see for example \cite{armeni2}). 
This fact shows that $A_{3}$ is negligible
for our purposes, as well as $A_{1}$ and $A_{2}$.

In conclusion, there is only one term that is relevant to our analysis, and
this is ${\rm G}_{6}$.

The most general bare lagrangian contains in the gravity sector the
following terms 
\begin{equation}
{\cal L}_{{\rm B}}=a_{{\rm B}}{\rm G}_{6}+b_{{\rm B}}RR_{\mu \nu }R^{\mu \nu
}+c_{{\rm B}}R^{3}+d_{{\rm B}}R\Box R+e_{{\rm B}}R_{\mu \nu }\Box R^{\mu \nu
},  \label{LB}
\end{equation}
where ${\rm G}_{6}$ (\ref{G6}) is continued to dimension $n$ as 
\[
-12R_{\mu \nu }R^{\nu \rho }R_{\rho }^{\mu }+{\frac{9n}{n-1}}RR_{\mu \nu
}R^{\mu \nu }-{\frac{3(n^{2}+4n-4)}{4(n-1)^{2}}}R^{3}+{\cal O}(W),
\]
and terms proportional to the Weyl tensor $W$, irrelevant to our analysis,
are omitted.

In the above list, there is one redundancy, as a reflection of the property
noted in (\ref{iden}) at the level of total derivatives. Indeed the
combination 
\begin{eqnarray}
&&-2(n-1)R_{\mu \nu }\Box R^{\mu \nu }-{\frac{n}{2}}(\nabla _{\mu }R)^{2}+{%
\frac{1}{2}}\Box R^{2}+(n-2)\nabla _{\mu }(R^{\mu \nu }\nabla _{\nu }R) 
\nonumber \\
&&+~2{\frac{n(n-1)}{n-2}}R_{\mu \nu }R^{\nu \rho }R_{\rho }^{\mu }-2{\frac{%
2n-1}{n-2}}RR_{\mu \nu }R^{\mu \nu }+{\frac{2}{n-2}}R^{3}  \label{ex}
\end{eqnarray}
is identically zero on conformally flat metrics, in arbitrary dimension $n$.
The derivative of the above expression (integrated over the space-time) with
respect to the conformal factor is $(n-6)$ times the same expression. For
this reason, it is negligible for our purposes.
Therefore we consistently remove the parameter $e_{{\rm B}}$ in (\ref{LB}).


In the end we remain with ${\rm G}_{6}$ and the three terms 
\begin{equation}
RR_{\mu \nu }R^{\mu \nu },~~~~~~~~~R^{3},~~~~~~~~~R\Box R,  \label{three}
\end{equation}
at the level of the bare lagrangian. All the other terms are either
proportional to the Weyl tensor or total derivatives in $n$ dimensions and
can therefore be omitted. The three terms (\ref{three})\ are forbidden at
criticality, since their integrals are not conformal-invariant in $n=6$.
Off-criticality, they do appear, but their coefficients, proportional to the
beta function, will be related to $a_{{\rm B}}$ in a way that we have to
uncover (and that should, in particular, agree with our prediction $\Delta
a=\Delta a^{\prime }$). There is some surviving ambiguity, actually, since $%
b_{{\rm B}}$, $c_{{\rm B}}$ and $d_{{\rm B}}$ can be shifted by arbitrary
additive constants.

The same three terms appear, multiplied by $(n-6)$, in the trace anomaly
equation. Moreover, four independent total-derivative terms, 
\begin{equation}
\Box (R_{\mu \nu }R^{\mu \nu }),~~~~~~~\Box R^{2},~~~~~~~~~\nabla _{\mu
}(R^{\mu \nu }\nabla _{\nu }R),~~~~~~~~~\Box ^{2}R,  \label{three2}
\end{equation}
appear at the level of the trace anomaly $\Theta $, with coefficients
related to the coefficients $a_{{\rm B}}$, $b_{{\rm B}}$, $c_{{\rm B}}$ and $%
d_{{\rm B}}$ (the coefficients of these terms in $\Theta $ do not come
multiplied by $(n-6)$).

It is worth recalling that in four dimensions we have the terms ${\rm G}_{4}$
and $R^{2}$ at the level of the bare lagrangian, ${\rm G}_{4}$, $R^{2}$ and $%
\Box R$ at the level of the trace anomaly. The integrability condition
requires that the coefficient of $R^{2}$ in the renormalized lagrangian be
proportional to the beta function and so vanish at criticality. The
coefficient of $\Box R$ in $\Theta $ is directly related to the coefficient of $R^{2}$%
.

\section{Perturbative calculation}

\setcounter{equation}{0}

In this section I apply Hathrell's techniques to the $\varphi ^{3}$-theory
in six dimensions, following the notation of \cite{hathrell2} (turned to the
Euclidean framework) as close as possible in order to facilitate the
reading. There are nevertheless unavoidable complications due to the large
number of invariants. I take the opportunity, in the derivation, to stress
the points that are related to the notion of pondered Euler density
introduced in this paper.

The complete bare lagrangian reads 
\[
{\cal L}_{{\rm B}}={\cal L}_{g}+{\cal L}_{\varphi },
\]
with
\begin{eqnarray*}
{\cal L}_{g} &=&a_{{\rm B}}{\rm G}_{6}+b_{{\rm B}}RR_{\mu \nu }R^{\mu \nu
}+c_{{\rm B}}R^{3}+d_{{\rm B}}R\Box R, \\
{\cal L}_{\varphi } &=&{\frac{1}{2}}(\nabla _{\mu }\varphi _{{\rm B}})^{2}+{%
\frac{1}{2}}\xi R\varphi _{{\rm B}}^{2}+{\frac{\lambda _{{\rm B}}}{3!}}%
\varphi _{{\rm B}}^{3}+{\frac{1}{2}}\eta _{{\rm B}}R\varphi _{{\rm B}}^{2},
\end{eqnarray*}
where $\xi =(n-2)/4(n-1)$ and $\eta $ is an independent coupling constant. A
flat-space lagrangian of the form (\ref{act}) does not have counterterms
linear in $\varphi $, due to the symmetry $(\varphi _{1},\varphi
_{2},\varphi _{3})\rightarrow (-\varphi _{1},-\varphi _{2},\varphi _{3})$.

The trace of the stress-tensor is 
\[
\Theta ={\frac{1}{\sqrt{g}}}{\frac{\delta S}{\delta \phi }}=\Theta
_{g}+\Theta _{\varphi }.
\]
Using the formulas reported in the Appendix, we find 
\begin{eqnarray}
\Theta _{g} &=&(n-6)\left[ a_{{\rm B}}{\rm G}_{6}+b_{{\rm B}}RR_{\mu \nu
}R^{\mu \nu }+c_{{\rm B}}R^{3}+d_{{\rm B}}R\Box R\right] -2(n-1)b_{{\rm B}%
}\Box (R_{\mu \nu }R^{\mu \nu })  \nonumber \\
&&\!\!\!\!\!\!\!\!\!\!\!\!\!\!\!{\ -\left[ {\frac{n+2}{2}}b_{{\rm B}%
}+6(n-1)c_{{\rm B}}+{\frac{n-2}{2}}d_{{\rm B}}\right] \Box R^{2}-2(n-2)b_{%
{\rm B}}\nabla _{\mu }(R^{\mu \nu }\nabla _{\nu }R)-4(n-1)d_{{\rm B}}\Box
^{2}R.}  \nonumber \\
\Theta _{\varphi } &=&-(n-6)\frac{1}{2}{\frac{\lambda _{{\rm B}}}{3!}}%
\varphi _{{\rm B}}^{3}-(n-1)\eta _{{\rm B}}\Box \varphi _{{\rm B}%
}^{2}+\left( {\frac{n}{2}}-1\right) {\rm E}_{{\rm B}},  \label{teta2}
\end{eqnarray}
where ${\rm E}_{{\rm B}}=[{\rm E}]={\frac{1}{\sqrt{g}}}\varphi _{{\rm B}}{%
\frac{\delta S_{\varphi }}{\delta \varphi _{{\rm B}}}}$ is the $\varphi _{%
{\rm B}}$-field equation; $\Theta _{\varphi }$ is the flat-space limit of
the trace operator. The coefficient $d$ plays the role of $a^{\prime }$ in
four dimensions \cite{athm}.

An indication in favour of the pondered Euler density is already visible, to
all orders in perturbation theory, in the expression of $\Theta $. Indeed,
we see that the four total-derivative terms of the kind (\ref{three2}) are
multiplied by three independent coefficients only. This integrability
condition must be compatible with the relationships between the coefficients
of the pondered Euler density. The crucial test is the ratio between the
factors of $\Box (R_{\mu \nu }R^{\mu \nu })$ and $\nabla _{\mu }(R^{\mu \nu
}\nabla _{\nu }R)$, which equals $5/4$ for $n=6$ in $\Theta $. The same
ratio appears in the expression (\ref{jeia}) of $J^{\alpha }.$ The remaining
factors can be checked only after lengthy work, which we now report on.

\subsection{The composite operator $\varphi^2$}

The renormalized field $\varphi $ is defined by 
\[
\varphi _{{\rm B}}=Z_{1}(\lambda ,n)\varphi .
\]
In flat space the renormalization of the composite operator $\varphi ^{2}$
reads 
\[
\lbrack \varphi ^{2}]=Z_{2}^{-1}(\lambda ,n)\varphi _{{\rm B}}^{2};
\]
$Z_{2}$ depends only on $\lambda $, because the other parameters (in
particular $\eta $) disappear when inserting $[\varphi ^{2}]$ in flat-space
correlators. The operator $\Box \varphi _{{\rm B}}^{2}$ is renormalized in
the same way as $\varphi _{{\rm B}}^{2}$. We can shift $\eta _{{\rm B}%
}\rightarrow \eta _{{\rm B}}+\eta Z_{2}^{-1}$, $\eta $ being an arbitrary
finite parameter, and still obtain finite results. Therefore we write 
\[
\eta _{{\rm B}}=(\eta +L_{\eta })Z_{2}^{-1},~~~~~~~~L_{\eta
}=\sum_{i=1}^{\infty }{\frac{\eta _{i}(\lambda )}{(n-6)^{i}}},
\]
where the pole part $L_{\eta }$ does not depend on $\eta $. In curved space
the most general expression for $[\varphi ^{2}]$ reads 
\begin{eqnarray}
\lbrack \varphi ^{2}] &=&Z_{2}^{-1}\varphi _{{\rm B}}^{2}+2\mu ^{n-6}\left[
(L_{\kappa ^{(1)}}+2\eta L_{\Lambda ^{(1)}})R_{\mu \nu }R^{\mu \nu }\right. 
\label{varf} \\
&&+(L_{\kappa ^{(2)}}+2\eta L_{\Lambda ^{(2)}}+3\eta ^{2}L_{\Sigma
^{(2)}})R^{2}\left. +(L_{\kappa ^{(3)}}+2\eta L_{\Lambda ^{(3)}})\Box
R\right] ,  \nonumber
\end{eqnarray}
where the pole series $L_{\kappa ,\Lambda ,\Sigma }$ have the same form as $%
L_{\eta }$. The $\eta $-dependences follow from the requirement of
finiteness of $\langle [\varphi ^{2}]\rangle $ in curved space. For example,
we observe that 
\begin{equation}
{\frac{1}{\sqrt{g}}}{\frac{\delta }{\delta \phi }}\langle [\varphi
^{2}]\rangle =-\langle \Theta \,[\varphi ^{2}]\rangle +{\frac{1}{\sqrt{g}}}%
\left\langle {\frac{\delta }{\delta \phi }}[\varphi ^{2}]\right\rangle =(%
{\rm finite}).  \label{*}
\end{equation}
We can take the flat-space limit and observe that $\left. \langle \Theta
\,[\varphi ^{2}]\rangle \right| $ is linear in $\eta $, which implies that
the coefficient of the curvature term $\Box R$ in $[\varphi ^{2}]$ is linear
in $\eta $. The terms $R_{\mu \nu }R^{\mu \nu }$ and $R^{2}$ contribute to
the second derivative of $\langle [\varphi ^{2}]\rangle $ with respect to
the conformal factor $\phi $, projected to flat space. Repeating the above
argument, we observe that $\left. \langle \Theta \,\Theta \,[\varphi
^{2}]\rangle \right| $ contains the term $\eta ^{2}~\langle [\varphi
^{2}]~\Box [\varphi ^{2}]~\Box [\varphi ^{2}]\rangle $. This term, however,
does not affect $R_{\mu \nu }R^{\mu \nu }$ and therefore the coefficient of $%
R_{\mu \nu }R^{\mu \nu }$ is linear in $\eta $, while the coefficient $R^{2}$
can be quadratic in $\eta $.

We proceed by observing that the insertion ${\frac{\partial S}{\partial \eta 
}}$ must produce finite results in correlators of elementary fields. We have 
\begin{eqnarray*}
{\frac{\partial S}{\partial \eta }} &=&{\frac{\partial S_{g}}{\partial \eta }%
}+{\frac{\partial S_{\varphi }}{\partial \eta }}, \\
{\frac{\partial S_{g}}{\partial \eta }} &=&\int \left[ {\frac{\partial a_{%
{\rm B}}}{\partial \eta }}{\rm G}_{6}+{\frac{\partial b_{{\rm B}}}{\partial
\eta }}RR_{\mu \nu }R^{\mu \nu }+{\frac{\partial c_{{\rm B}}}{\partial \eta }%
}R^{3}+{\frac{\partial d_{{\rm B}}}{\partial \eta }}R\Box R\right] , \\
{\frac{\partial S_{\varphi }}{\partial \eta }} &=&\int {\frac{1}{2}}%
\,R\,\varphi _{{\rm B}}^{2}\,Z_{2}^{-1}.
\end{eqnarray*}
We can therefore write, by using (\ref{varf}) and equating the various
independent pole terms, 
\begin{eqnarray*}
a_{{\rm B}} &=&\mu
^{n-6}(a+L_{a}),~~~~~~~~~~~~~~~~~~~~~~~~~~~~~~~~~~~~~~~~~b_{{\rm B}}=\mu
^{n-6}(b+L_{b}+\eta L_{\kappa ^{(1)}}+\eta ^{2}L_{\Lambda ^{(1)}}), \\
c_{{\rm B}} &=&\mu ^{n-6}(c+L_{c}+\eta L_{\kappa ^{(2)}}+\eta ^{2}L_{\Lambda
^{(2)}}+\eta ^{3}L_{\Sigma ^{(2)}}),~~~d_{{\rm B}}=\mu ^{n-6}(d+L_{d}+\eta
L_{\kappa ^{(3)}}+\eta ^{2}L_{\Lambda ^{(3)}}),
\end{eqnarray*}
$a$, $b$, $c$ and $d$ denoting finite independent parameters, while $%
L_{a,b,c,d}$ etc. are pole series in $\lambda $ of the same form as $L_{\eta
}$.

\subsection{The renormalization-group equations}

Finite and pole parts are related by the condition 
\begin{equation}
\mu{\frac{{\rm d}}{{\rm d}\mu}}{\cal L}_{{\rm B}}=0,  \label{bare}
\end{equation}

In particular, the ambiguities related to the addition of arbitrary finite
local terms (\ref{three}) to the action are almost completely fixed. Finite
local terms add, unchanged, to the one-particle irreducible effective action 
$\Gamma $ and affect $\Theta $ via the total derivatives of the list (\ref
{three2}). This kind of ambiguity does not concern $a_{{\rm B}}$, since the $%
\phi $-variation of $\int {\rm G}_{6}$ is zero in six dimensions (see the
Appendix). Moreover, these additions have to preserve (\ref{bare}) and
therefore cannot depend on the renormalized coupling constant $\lambda $.
More precisely, if $b_{{\rm B}}\rightarrow b_{{\rm B}}+b^{\prime }\mu ^{n-6}$
with $b^{\prime }$ finite, we have in the limit $n=6$, $\mu {\frac{{\rm d}%
b^{\prime }}{{\rm d}\mu }}=\beta {\frac{\partial b^{\prime }}{\partial
\lambda }}=0$. The shift $b^{\prime }$ can thus depend on the coupling
constant only in families of conformal field theories ($\beta =0$), such as
N=4 supersymmetric Yang--Mills theory. This case, however, does not concern
us here. Similar conclusions apply to $c_{{\rm B}}$ and $d_{{\rm B}}$.

Let us start from $D\eta _{{\rm B}}=0$, where 
\[
D=\mu {\frac{{\rm d}}{{\rm d}\mu }}={\frac{\partial }{\partial \mu }}+\hat{%
\beta}(\lambda ){\frac{\partial }{\partial \lambda }}+\mu {\frac{{\rm d}\eta 
}{{\rm d}\mu }}{\frac{\partial }{\partial \eta }}+\cdots 
\]
Here $\hat{\beta}=\mu {\frac{{\rm d}\lambda }{{\rm d}\mu }=}\frac{1}{2}%
(n-6)\lambda +\beta $ is the $n$-dimensional beta-function and $\beta =-D\ln
Z_{3}$ ($\lambda _{{\rm B}}=\lambda Z_{3}(\lambda ,n)\mu ^{-(n-6)/2}$) is
the six-dimensional one. We write 
\[
(D+\gamma )Z_{1}^{-1}=0,~~~~~~~~~~~(D+\delta )Z_{2}^{-1}=0,
\]
$\gamma $ and $\delta $ being finite functions, the anomalous dimensions of $%
\varphi $ and $\varphi ^{2}$, respectively. Calling $\beta _{\eta }(\lambda )
$ the finite function $(D-\delta )\eta =~\mu {\frac{{\rm d}\eta }{{\rm d}\mu 
}}~-\delta \eta $, we can write the renormalization-group equation for the $%
\eta $-poles in the form 
\begin{equation}
(D-\delta )L_{\eta }=-\beta _{\eta }=\frac{\lambda }{2}{\frac{\partial }{%
\partial \lambda }}\eta _{1}(\lambda ),  \label{eta1}
\end{equation}
the last equality being obtained by writing the finite term in $(D-\delta
)L_{\eta }$ explicitly.

Similarly, $Da_{{\rm B}}=Db_{{\rm B}}=Dc_{{\rm B}}=Dd_{{\rm B}}=0$, give the
other renormalization-group equations. Calling $\beta _{a,b,c,d}(\lambda )$
etc. the various finite parts, functions of the coupling $\lambda $, we have 
\begin{eqnarray}
Da &=&\hat{\beta}_{a}  \nonumber \\
D(b,c,d) &=&\hat{\beta}_{b,c,d}+\eta \beta _{\kappa ^{(1,2,3)}}+\eta
^{2}\beta _{\Lambda ^{(1,2,3)}}+\eta ^{3}\beta _{\Sigma ^{(2)}},  \nonumber
\\
\hat{\beta}_{a,b,c,d} &=&-(n-6)(a,b,c,d)+\beta _{a,b,c,d}(\lambda ), 
\nonumber \\
\lbrack (n-6)+D]L_{a} &=&-\beta _{a}(\lambda )=\frac{1}{2\lambda }{\frac{%
\partial }{\partial \lambda }}\left( \lambda ^{2}a_{1}\right) ,  \nonumber \\
\lbrack (n-6)+D]L_{b,c,d}+\beta _{\eta }L_{\kappa ^{(1,2,3)}} &=&-\beta
_{b,c,d}=\frac{1}{2\lambda }{\frac{\partial }{\partial \lambda }}\left(
\lambda ^{2}(b,c,d)_{1}\right) ,  \nonumber \\
\lbrack (n-6)+D+\delta ]L_{\kappa ^{(1,2,3)}}+2\beta _{\eta }L_{\Lambda
^{(1,2,3)}} &=&-\beta _{\kappa ^{(1,2,3)}}=\frac{1}{2\lambda }{\frac{%
\partial }{\partial \lambda }}\left( \lambda ^{2}\kappa
_{1}^{(1,2,3)}\right) ,  \label{penu} \\
\lbrack (n-6)+D+2\delta ]L_{\Lambda ^{(1,2,3)}}+3\beta _{\eta }L_{\Sigma
^{(2)}} &=&-\beta _{\Lambda ^{(1,2,3)}}=\frac{1}{2\lambda }{\frac{\partial }{%
\partial \lambda }}\left( \lambda ^{2}\Lambda _{1}^{(1,2,3)}\right) , 
\nonumber \\
\lbrack (n-6)+D+3\delta ]L_{\Sigma ^{(2)}} &=&-\beta _{\Sigma ^{(2)}}=\frac{1%
}{2\lambda }{\frac{\partial }{\partial \lambda }}\left( \lambda ^{2}\Sigma
_{1}^{(2)}\right) .  \nonumber
\end{eqnarray}
We have compressed the formulas in a self-evident notation. We recall that $%
a_{1}$, $b_{1}$, etc., are the simple poles of the various counterterms.

At a first reading, the reader can follow the various steps focusing on the
quantity $a$. The rest is entirely analogous, with the complication due to
the renormalization mixing with the operator $\varphi ^{2}$.

\subsection{The composite operator $\varphi^3$}

The analysis of the composite operator $\varphi ^{2}$ has to be repeated for 
$\varphi ^{3}$, by studying the insertions of ${\frac{%
\partial S}{\partial \lambda }}$ in correlators. Actually, this procedure
produces the integral of the renormalized operator $[\varphi ^{3}]$, rather
than $[\varphi ^{3}]$ itself. Therefore $[\varphi ^{3}]$ is obtained up to
arbitrary total-derivative terms of the form (\ref{three2}) and $\Box [\varphi ^{2}]$,
multiplied by pole series that we write as 
\begin{equation}
{\frac{1}{n-6}}\left( f^{(1,2,3,4)}+L_{f^{(1,2,3,4)}}+\eta
(g^{(1,2,3,4)}+L_{g^{(1,2,3,4)}})+\cdots \right) ~~~~~~~~~{\rm and}~~~~~~~~~~%
{\frac{1}{n-6}}(e+L_{e}),  \label{ho}
\end{equation}
respectively, $f^{(1,2,3,4)}$, $g^{(1,2,3,4)},$ etc., and $e$ denoting
finite functions. The result for ${\frac{1}{3!}}\mu ^{-(n-6)/2}[\varphi ^{3}]
$, as well as details of the calculation, are given in the appendix, formula
(\ref{a1}).

At this point, using (\ref{teta2}) we obtain the manifestly finite formula
for the trace anomaly, which is 
\begin{eqnarray}
\Theta  &=&-{\frac{\hat{\beta}}{3!}}[\varphi ^{3}]\mu ^{-(n-6)/2}+\left( {%
\frac{n}{2}}-1+\gamma \right) [{\rm E}]-(e+(n-1)\eta )\Box [\varphi ^{2}]-{%
\frac{1}{2}}(\beta _{\eta }+\eta \delta )R[\varphi ^{2}]  \nonumber \\
&&-~\mu ^{n-6}\left[ \hat{\beta}_{a}{\rm G}_{6}+(\hat{\beta}_{b,c,d}+\eta
\beta _{\kappa ^{(1,2,3)}}+\eta ^{2}\beta _{\Lambda ^{(1,2,3)}}+\eta
^{3}\beta _{\Sigma ^{(2)}})(RR_{\mu \nu }R^{\mu \nu },R^{3},R\Box R)\right] 
\nonumber \\
&&-~\mu ^{n-6}\left[ (f^{(1)}+2(n-1)b+\eta g^{(1)}+\eta ^{2}h^{(1)})\Box
R_{\mu \nu }R^{\mu \nu }\right.   \label{tttt} \\
&&+\left( f^{(2)}+\frac{1}{2}(n+2)b+6(n-1)c+{\frac{1}{2}}(n-2)d+\eta
g^{(2)}+\eta ^{2}h^{(2)}\right) \Box R^{2}  \nonumber \\
&&+~(f^{(3)}+4(n-1)d+\eta g^{(3)})\Box ^{2}R+\left. (f^{(4)}+2(n-2)b+\eta
g^{(4)}+\eta ^{2}h^{(4)})\nabla _{\mu }(R^{\mu \nu }\nabla _{\nu }R)\right] 
\nonumber
\end{eqnarray}
and the consistency relations (\ref{a2}) implied by finiteness. Other
remarkable relations are obtained by applying the operator $\hat{\beta}^{-1}D%
\hat{\beta}$ to $[\varphi ^{3}]$. The operator $\hat{\beta}^{-1}D\hat{\beta}$
has the property that it produces finite quantities when acting on finite
quantities. Re-expressing $\hat{\beta}[\varphi ^{3}]$ in terms of $\Theta $
and using $D\Theta =0$, finiteness implies the important relations (\ref
{appul}).

We are now ready to study the lowest-order contributions to some functions.
The anomalous dimension $\delta $ of $[\varphi ^{2}]$ is clearly ${\cal O}%
(\lambda ^{2})$. The quantity $e/\lambda $ is the lowest-order pole of the
mixing between $[\varphi ^{3}]$ and $\Box [\varphi ^{2}]$, and is ${\cal O}%
(\lambda ^{3})$. Therefore $e={\cal O}(\lambda ^{4})$. The first equation of
the list (\ref{appul}) gives $\beta _{\eta }={\cal O}(\lambda ^{6})$.
Formula (\ref{eta1}) gives $\eta _{1}(\lambda )={\cal O}(\lambda ^{6})$ also.

According to the definition (\ref{a1}) of $[\varphi ^{3}]$, the quantity $%
f^{(3)}/\lambda $ is the first pole of the mixing between $[\varphi ^{3}]$
and $\Theta $ ({\sl via} $\Box ^{2}R$). In particular, 
\[
({\rm finite})={\frac{1}{\sqrt{g}}}{\frac{\delta }{\delta \phi }}\langle
[\varphi ^{3}]\rangle =-\langle \Theta [\varphi ^{3}]\rangle +\left\langle {%
\frac{1}{\sqrt{g}}}{\frac{\delta }{\delta \phi }}[\varphi ^{3}]\right\rangle
.
\]
Projecting this equation to flat space, we see that the pole $%
f^{(3)}/\lambda $ is cancelled by the first pole of $\langle \Theta [\varphi
^{3}]\rangle $. This is at least ${\cal O}(\lambda ^{3})$, as we now show,
so that we expect $f^{(3)}={\cal O}(\lambda ^{4})$.

In flat space, the operator $\Theta (x)$ reads 
\[
\left. \Theta \right| =-{\frac{\hat{\beta}}{3!}}[\varphi ^{3}]\mu
^{-(n-6)/2}+\left( {\frac{n}{2}}-1+\gamma \right) [{\rm E}]-(e+(n-1)\eta
)\Box [\varphi ^{2}].
\]
We recall that $\beta ={\cal O}(\lambda ^{3}),$ while $e={\cal O}(\lambda
^{4}).$ Moreover, we can write $\hat{\beta}=\frac{1}{2}(n-6)\lambda +\beta =%
{\cal O}(\lambda ^{3})$. Indeed, the term proportional to $(n-6)$ selects a
higher-loop singularity, and we can formally say that $(n-6)$ is ${\cal O}%
(\lambda ^{2})$. Finally, $\gamma ={\cal O}(\lambda ^{2})$, but we can
neglect the term in $[{\rm E}]$ in $\langle \left. \widetilde{\Theta }%
\right| [\varphi ^{3}]\rangle $, since a simple functional argument shows
that $\langle [{\rm E}]~P\rangle =0$ for any operator $P.$ We have thus
proved that $f^{(3)}={\cal O}(\lambda ^{4})$. There is, however, a
cancellation between the two-loop contribution singled out by the term ${%
\frac{\lambda }{2}}(n-6)$ in $\hat{\beta}$ and the one-loop contribution
coming from $\beta $. This enhancement, which produces the result that $%
f^{(3)}$ is actually ${\cal O}(\lambda ^{6})$, will be proved in the next
section. Precisely, it will be proved that $\beta _{d}={\cal O}(\lambda ^{8})
$. Then the claimed result follows from the fifth formula from the bottom of
the list (\ref{appul}).

We now consider the quantity $g^{(3)}/\lambda $. We can repeat the above
analysis and consider the $\eta $-term in $\left. \Theta \right| $. The pole 
$g^{(3)}/\lambda $ cancels the first pole of $\langle [\varphi ^{3}]\,\Box
[\varphi ]^{2}\rangle $, which is ${\cal O}(\lambda )$. So, $g^{(3)}={\cal O}%
(\lambda ^{2})$. No enhancement takes place here.

Moreover, the fourth formula of (\ref{appul}) from the bottom implies that $%
\beta _{\kappa ^{(3)}}={\cal O}(\lambda ^{4})$. The same result can be
proved, using formula (\ref{*}) projected to flat space and the definition (%
\ref{varf}) of $[\varphi ^{2}]$: $\kappa _{1}^{(3)}$ is the first-loop pole
of $\langle \beta [\varphi ^{3}]\,[\varphi ^{2}]\rangle $. Taking the $\eta $%
-term, a similar argument proves that $\beta _{\Lambda ^{(3)}}={\cal O}(1)$.
Similar considerations show that $f^{(1,2,4)}={\cal O}(\lambda ^{6})$, $%
g^{(1,2,4)}={\cal O}(\lambda ^{2})$, $\beta _{\kappa ^{(1,2)}}={\cal O}%
(\lambda ^{4})$, $\beta _{\Lambda ^{(1,2)}}={\cal O}(1)$.


Before closing this section, let us discuss the parameter $\eta $ in detail,
in particular the renormalization-group equation $D\eta =\delta \eta +\beta
_{\eta }.$ Writing, $\eta =u(\lambda )\eta ^{\prime }+v(\lambda ),$ we have $%
D\eta ^{\prime }=0,$ and for $n=6,$ $u(\lambda )\sim \lambda ^{\delta
_{1}/\beta _{1}}$ (where $\delta =\delta _{1}\lambda ^{2}+{\cal O}(\lambda
^{4}),$ $\beta =\beta _{1}\lambda ^{3}+{\cal O}(\lambda ^{5})$) and $%
v(\lambda )={\cal O}(\lambda ^{4}).$ The parameter $\eta ^{\prime }$ is the
true independent coupling constant of the term $R\varphi ^{2}$ in the
lagrangian and can be set to zero. A non-zero value of this parameter would
produce non-integer powers of $\lambda $ and therefore the choice $\eta
^{\prime }=0$ is universal. This can be better apppreciated by observing
that the solution for $u(\lambda )$ reads 
\[
u(\lambda )=\exp \left( \int_{\lambda _{0}}^{\lambda }\frac{\delta \left(
\lambda ^{\prime }\right) }{\beta (\lambda ^{\prime })}{\rm d}\lambda
^{\prime }\right) .
\]
The arbitrariness of $\lambda _{0}$ is the same as the arbitrariness of $%
\eta ^{\prime }$. Now, $u(\lambda )$ is not proportional to the beta
function, so $\Theta $ contains a term $\eta ^{\prime }u(\lambda )\Box
[\varphi ^{2}]$. To restore scale invariance at the fixed points, $\Theta $
should be proportional to the beta function, up to field equations.
Therefore we have to set $\eta ^{\prime }=0.$ The remaining $\Box [\varphi
^{2}]$-term is indeed proportional to the beta function, since using the
first relation of (\ref{appul}) the coefficient of the $\Box [\varphi ^{2}]$%
-term in $\Theta $ reads 
\[
e+(n-1)\eta =(n-1)\frac{\hat{\beta}}{\delta }\frac{\partial \eta }{\partial
\lambda }.
\]
Observe that $\frac{1}{\delta }\frac{\partial \eta }{\partial \lambda }$ is
regular also in the free-field limit. For a generic interacting fixed point, 
$\delta _{*}>0$. We conclude that only the choice $\eta ^{\prime }=0$ is
consistent with the RG\ interpolation problem. At a criticality we have,
correctly, 
\[
\left. \Theta \right| =\left( {\frac{n}{2}}-1+\gamma _{*}\right) [{\rm E}]~,
\]
$\gamma _{*}$ being the anomalous dimension of the elementary field $\varphi 
$.

\subsection{The $\Theta $ two-, three- and four-point function}

We now explain the basic reason why the pondered Euler density comes out
naturally of the calculation. It is convenient to define 
\[
\widetilde{\Theta }=\sqrt{g}\Theta =\frac{\delta S}{\delta \phi }.
\]
We study the first, second and third derivatives of $\tilde{\Theta}$, which
are related to the two-, three- and fourth-point functions of $\widetilde{%
\Theta }$ itself. The derivatives of a finite quantity, like $\langle \tilde{%
\Theta}\rangle $, are finite. Therefore we have, for example, 
\begin{equation}
{\frac{\delta \langle \widetilde{\Theta }\rangle }{\delta \phi }}=-\langle 
\widetilde{\Theta }~\widetilde{\Theta }\rangle +\left\langle {\frac{\delta }{%
\delta \phi }}\tilde{\Theta}\right\rangle =({\rm finite}).  \label{twop}
\end{equation}
First, we can work out relations between the poles. Then, using the RG
equations, we can obtain information about the finite parts.

The two-point function fixes the order of the first radiative correction to
the coefficient $d$, in particular the coefficient $d_{1}$ of the first-pole
term.

Let us define 
\[
\bar{\widetilde{\Theta }}=\widetilde{\Theta }-\left( \frac{n}{2}-1\right) 
\sqrt{g}{\rm E}=\widetilde{\Theta }-\left( \frac{n}{2}-1\right) {\rm \tilde{E%
}},
\]
so that equation (\ref{twop}) can be re-expressed as  
\begin{equation}
-\langle \bar{\widetilde{\Theta }}~\bar{\widetilde{\Theta }}\rangle
+\left\langle {\frac{\delta }{\delta \phi }}\tilde{\Theta}\right\rangle =(%
{\rm finite}).  \label{twp2}
\end{equation}
We can neglect the term in $[{\rm E}]$ in $\langle \widetilde{\Theta }~%
\tilde{\Theta}\rangle $, since $%
\langle [{\rm E}]~P\rangle =0$ for any $P.$ We have 
\[
\left. \bar{\widetilde{\Theta }}\right| =\frac{n-6}{2}%
\frac{\lambda _{{\rm B}}}{3!}\varphi _{{\rm B}}^{3}-(n-1)\eta _{{\rm B}}\Box
\varphi _{{\rm B}}^{2}=-{\frac{\hat{\beta}}{3!}}[\varphi ^{3}]\mu
^{-(n-6)/2}+\gamma [{\rm E}]-(e+(n-1)\eta )\Box [\varphi ^{2}].
\]
We can write in general 
\[
\left. \bar{\widetilde{\Theta }}\right| =\hat{\beta}%
Q,\qquad Q=-{\frac{1}{3!}}[\varphi ^{3}]\mu ^{-(n-6)/2}+\frac{\gamma }{\hat{%
\beta}}[{\rm E}]-(n-1)\frac{1}{\delta }\frac{\partial \eta }{\partial
\lambda }\Box [\varphi ^{2}].
\]
The operator $Q$ is particularly useful. It contains poles, but we will see
that they are harmless, since they multiply just the field equations. An
important property of $Q$ is $D(\hat{\beta}Q)=0$ in flat space.

We recall that $\beta ,\hat{\beta}={\cal O}(\lambda ^{3}),$ while $e,\eta =%
{\cal O}(\lambda ^{4}),$ $\gamma ={\cal O}(\lambda ^{2})$. Naively, we
expect $d_{1}={\cal O}(\lambda ^{6}).$ We now prove that there is an
enhancement due to the RG equations and $d_{1}={\cal O}(\lambda ^{8})$. This
phenomenon is similar to a phenomenon occurring in four dimensions \cite
{hathrell2}, but here is implies that the first radiative corrections to the
quantity $a$ are four-loop, rather than three-loop.
We write the poles of the two-point function as 
\begin{equation}
\langle Q(x)~Q(0)\rangle =\mu ^{n-6}\left( \frac{n-6}{\hat{\beta}}\right)
^{2}L_{X}~\Box ^{3}\delta (x)+({\rm finite}),  \label{apoi}
\end{equation}
where $L_{X}$ is a pole series of the usual form. The divergent terms are local,
since subdivergences are originated 
only from vertex graphs of elementary fields and at most one insertion of $Q$. 
Therefore they are automatically subtracted away.

We see that the first pole 
$X_{1}$ is, by definition, ${\cal O}(\lambda ^{2})$. Writing $%
X_{1}=x_{1}\lambda ^{2}$ we find by explicit computation 
\[
x_{1}=-\frac{1}{4\cdot 6!}\frac{1}{(4\pi )^{6}}.
\]

Formula (\ref{twp2}) gives 
\[
\widetilde{d}_{1}=\frac{X_{3}}{200},
\]
where $X_{3}$ is the triple pole of $L_{X}$. We put a tilde on $d_{1}$ to
stress that the first pole in consideration does not come just from $L_{d},$
but from the full sum $L_{d}+\eta L_{\kappa ^{(3)}}+\eta ^{2}L_{\Lambda
^{(3)}}$ appearing in the definition of $d_{{\rm B}}$ given in section 4.1.
The $\eta $-contributions are indeed ${\cal O}(\lambda ^{8})$ also.

We can compute $X_{3}$ as follows. We apply $\hat{\beta}^{-2}D\hat{\beta}^{2}
$ to (\ref{apoi}) and observe that $\hat{\beta}^{-2}D\hat{\beta}^{2}({\rm %
finite})=({\rm finite}).$ Moreover, we recall that $D(\hat{\beta}Q)=0.$ We
have, therefore:
\begin{equation}
\left( \frac{n-6}{\hat{\beta}}\right) ^{2}[(n-6)+D]L_{X}=({\rm finite})=%
\frac{2}{\lambda ^{3}}\frac{\partial }{\partial \lambda }\left( \lambda
^{2}X_{1}\right) .  \label{erf}
\end{equation}
The right-hand side is obtained by writing the unique finite contribution to
the left-hand side. Solving the recursion relations we arrive at 
\[
\frac{\partial }{\partial \lambda }\left( \lambda ^{2}X_{3}\right) =-8\frac{%
\beta }{\lambda ^{2}}\int^{\lambda }{\rm d}\lambda ^{\prime }~\lambda
^{\prime 4}X_{1}(\lambda ^{\prime })\frac{\partial }{\partial \lambda
^{\prime }}\left( \frac{\beta (\lambda ^{\prime })}{\lambda ^{\prime 3}}%
\right) ,
\]
and therefore 
\begin{equation}
X_{3}=-\frac{1}{5}\beta _{1}\beta _{2}x_{1}\lambda ^{8},\qquad \widetilde{d}%
_{1}=\frac{1}{4000\cdot 6!}\beta _{1}\beta _{2}\frac{\lambda ^{8}}{(4\pi
)^{6}}.  \label{puppo}
\end{equation}

We now study the second derivative of $\langle \widetilde{\Theta }(x)\rangle 
$ with respect to $\phi $, 
\begin{equation}
\langle \widetilde{\Theta }(x)~\widetilde{\Theta }(y)~\widetilde{\Theta }%
(z)\rangle -\left\langle {\frac{\delta \widetilde{\Theta }(x)}{\delta \phi
(y)}}~\widetilde{\Theta }(z)\right\rangle -\left\langle {\frac{\delta 
\widetilde{\Theta }(x)}{\delta \phi (z)}}~\widetilde{\Theta }%
(y)\right\rangle -\left\langle {\frac{\delta \widetilde{\Theta }(y)}{\delta
\phi (z)}}~\widetilde{\Theta }(x)\right\rangle +\left\langle {\frac{\delta
^{2}\widetilde{\Theta }(x)}{\delta \phi (y)\delta \phi (z)}}\right\rangle =(%
{\rm finite}).
\label{form}
\end{equation}
The $\tilde\Theta$ three-point function contains non-local divergences associated with
vertices with two $\tilde\Theta$-legs in the same point. These, however, are subtracted away
by the middle terms of (\ref{form}) and the remaining overall divergence (the last term on the left-hand
side of (\ref{form})) is indeed purely local.
(\ref{form}) can be written in the schematic form 
\begin{equation}
\langle \bar{\widetilde{\Theta }}~\bar{\widetilde{\Theta }}~\bar{\widetilde{%
\Theta }}\rangle -3\left\langle \bar{\widetilde{\Theta }}~\frac{\Delta \bar{%
\widetilde{\Theta }}}{\Delta \phi }\right\rangle +\left\langle {\frac{\delta
^{2}\tilde{\Theta}(x)}{\delta \phi ^{2}}}\right\rangle =({\rm finite}),
\label{oi}
\end{equation}
where we have defined 
\[
\frac{\Delta \bar{\widetilde{\Theta }}}{\Delta \phi }={\frac{\delta \bar{%
\widetilde{\Theta }}}{\delta \phi }-}\frac{n-2}{2}\varphi _{{\rm B}}{\frac{%
\delta \bar{\widetilde{\Theta }}}{\delta \varphi _{{\rm B}}}}
\]
and used 
\begin{equation}
\langle [\tilde{{\rm E}}]~A~B\rangle =\left\langle \varphi _{{\rm B}}{\frac{%
\delta A}{\delta \varphi _{{\rm B}}}}~B\right\rangle +\left\langle A~\varphi
_{{\rm B}}{\frac{\delta B}{\delta \varphi _{{\rm B}}}}\right\rangle .
\label{burp}
\end{equation}
We write
\[
\left.\frac{\Delta \bar{\widetilde{\Theta }}(x)}{\Delta \phi(y) }\right|=\hat{\beta}%
\frac{\Delta Q(x)}{\Delta \phi (y)}+8(n-1)^2d_{\rm B}\Box^3\delta(x-y).
\]
The $d_{\rm B}$-term here originates disconnected contributions to the correlators.
There is no such contribution in (\ref{oi}), but there will be some in the four-point function.
Moreover, we write
\begin{equation}
\langle Q~Q~Q\rangle_* -\frac{3}{\hat{\beta}}\langle Q~\frac{\Delta Q}{\Delta
\phi }\rangle_* =\mu ^{n-6}\left( \frac{n-6}{\hat{\beta}}\right)
^{3}L_{Y}\times \ {\rm local\ structure}+({\rm finite}) .
\label{corrode}
\end{equation}
Here there are also non-local divergent terms coming from
subdivergences with a vertex of two $Q$-legs in the same point. Not all of them are 
automatically subtracted by the second term of (\ref{corrode}). 
The star subscript means that we subtract them away by hand.

The local structure of formula (\ref{corrode}) is a sum of expressions
of the form $\partial ^{p}\delta (x-y)~\partial ^{q}\delta (x-z),$
with $p+q=6$, $p,q\geq 2$, fixed by the term $\left. \langle {\frac{%
\delta ^{2}\tilde{\Theta}(x)}{\delta \phi ^{2}}}\rangle \right| $ in (\ref
{oi}). 

Applying $\hat{\beta}^{-3}D\hat{\beta}^{3}$ and using $D
\frac{\Delta \bar{\widetilde{\Theta }}}{\Delta \phi }=0$ we can prove an RG
equation similar to (\ref{erf}) and show that there is an enhancement again.
This would be straightforward in the absence of the subtractions just mentioned.
The result is that one finds an equation of the type (\ref{erf}), but with 
a non-zero right-hand side. It is easy to see, however, by direct inspection
of the structure and graphs of (\ref{corrode}), that these corrections are of higher order
(${\cal O}(\lambda^4)$ for $Y_1$, ${\cal O}(\lambda^6)$ for $Y_2$, etc. up to ${\cal O}(\lambda^{10})$ for $Y_4$) and we do not
need any enhancement to discard these. This fact is actually natural, since subdivergences
cannot contribute to the lowest orders, which is all what we are interested in here.
In four dimensions the matter is entirely similar \cite{hathrell2,hathrell}.
We also remark, without giving further details, 
that there is one peculiar local structure which is not affected by non-local subdivergences.
This is not sufficient, however, for us, since we need two conditions. For this reason we are compelled 
to inspect the
subdivergences originated 
by the vertices with two $Q$-legs in the same point. The matter is simpler
for the four-point function, where there is one special
local structure and we need precisely one condition.

A direct inspection of (\ref{corrode}) proves that to the lowest order $Y_{1}={\cal %
O}(\lambda ^{2})$, so that the RG equations imply $Y_{2}={\cal %
O}(\lambda ^{4})$ and $Y_{3}={\cal %
O}(\lambda ^{6})$ and one
 naively expects $Y_{4}={\cal O}(\lambda ^{8})$.
The enhancement fixes instead $Y_{4}={\cal O}(\lambda
^{10}).$

To help studying $Y_1$ we observe that 
\[
\frac{\Delta Q}{\Delta \phi }{=}\frac{1}{2}(6-n)Q+\eta\-{\rm terms}.
\]
The $\eta $-terms are of order ${\cal O}(\lambda )$, at least, quadratic in $%
\varphi $, and also contain poles. The important fact is that $\Delta Q/{%
\Delta \phi }$ is of higher order with respect to $Q$. Finally, in $Q$ the
field equation ${\rm E}$ is multiplied by $\gamma /\hat{\beta},$ which is
formally ${\cal O}(\lambda /(n-6))$. Let us note that $\langle [{\rm E}][{\rm E}][{\rm E}%
]\rangle =0$, which follows from (\ref{burp}).

Now, $Y_1={\cal O}(\lambda^2)$ means a simple pole in (\ref{corrode}) of order
$1/\lambda$. This can be originated, indeed, by the denominator $1/\hat\beta$, but
it is easy to check that contributions worse tan this (finite or divergent,
local or non-local) simply do not exist in (\ref{corrode}).

We conclude the analysis of the three-point function by stating that 
\begin{equation}
\left. \left\langle {\frac{\delta ^{2}\widetilde{\Theta }(x)}{\delta \phi
(y)\delta \phi (z)}}\right\rangle \right| =({\rm finite}),  \label{oi2}
\end{equation}
to the order $\lambda ^{8}$ included. 
In particular, we study the simple poles of (%
\ref{oi2}) and set them to zero. These are linear combinations of $b_{1},$ 
$c_{1}$ and $d_{1}$ (see below). Because of the one-to-one correspondence
between pole terms and finite terms, it is like setting the finite part of $%
\left. {\frac{\delta ^{2}\widetilde{\Theta }(x)}{\delta \phi (y)\delta \phi
(z)}}\right| $ to zero and this is the first step for the definition of the
pondered Euler density.

Then, the argument is repeated for the four-point function, first writing 
\begin{equation}
-\langle \bar{\widetilde{\Theta }}~\bar{\widetilde{\Theta }}~\bar{\widetilde{%
\Theta }}~\bar{\widetilde{\Theta }}\rangle -3\left\langle \frac{\Delta \bar{%
\widetilde{\Theta }}}{\Delta \phi }~\frac{\Delta \bar{\widetilde{\Theta}}}{\Delta \phi 
}\right\rangle -4\left\langle \bar{\widetilde{\Theta }}~\frac{\Delta ^{2}%
\bar{\widetilde{\Theta }}}{\Delta \phi ^{2}}\right\rangle +6\left\langle 
\bar{\widetilde{\Theta }}~\bar{\widetilde{\Theta }}~\frac{\Delta \bar{\widetilde{%
\Theta}}}{\Delta \phi }\right\rangle +\left\langle {\frac{\delta ^{3}%
\widetilde{\Theta }(x)}{\delta \phi ^{3}}}\right\rangle =({\rm finite}),
\label{hugh}
\end{equation}
and then observing that this formula reduces to 
\begin{equation}
\left. \left\langle {\frac{\delta ^{3}\widetilde{\Theta }(x)}{\delta \phi
(y)\delta \phi (z)\delta \phi (w)}}\right\rangle \right| =({\rm finite}),
\label{con}
\end{equation}
to the order we are interested in. This result can be obtained following 
the same 
procedure that we have applied for the two- and three-point functions
and keeping into account that, despite there
are disconnected contributions, 
they sum up to finite terms and can therefore be neglected throughout. 
Indeed, the disconnected contributions to
(\ref{hugh}) read in flat space
\[
-3\langle \bar{\widetilde{\Theta }}~\bar{\widetilde{\Theta }}\rangle\langle\bar{\widetilde{%
\Theta }}~\bar{\widetilde{\Theta }}\rangle -3\left\langle \frac{\Delta \bar{%
\widetilde{\Theta }}}{\Delta \phi }\right\rangle\left\langle\frac{\Delta \bar{\widetilde{\Theta}}}{\Delta \phi 
}\right\rangle +6\langle
\bar{\widetilde{\Theta }}~\bar{\widetilde{\Theta }}\rangle~\left\langle \frac{\Delta \bar{\widetilde{%
\Theta}}}{\Delta \phi }\right\rangle =-3\left(\langle \bar{\widetilde{\Theta }}~\bar{\widetilde{\Theta }}\rangle
-\left\langle \frac{\Delta \bar{%
\widetilde{\Theta }}}{\Delta \phi }\right\rangle\right)^2=({\rm finite}),
\]
using (\ref{twp2}).

We study the order-$\lambda ^{8}$ simple poles of (\ref{oi2}) and (\ref{con}%
). The relevant contribution to $\widetilde{\Theta }$ is 
\begin{equation}
\frac{1}{n-6}\left[ a_{2}\tilde{{\rm G}}_{6}-10\widetilde{b}_{1}\Box \left(
R_{\mu \nu }R^{\mu \nu }\right) -8\widetilde{b}_{1}\nabla _{\mu }\left(
R^{\mu \nu }\nabla _{\nu }R\right) -(30\widetilde{c}_{1}+4\widetilde{b}_{1}+2%
\widetilde{d}_{1})\Box R^{2}-20\widetilde{d}_{1}\Box ^{2}R\right] .
\label{auto}
\end{equation}
We recall that the tildes on $b_{1},c_{1},d_{1}$ denote the full first poles 
of $b_{{\rm B}},c_{{\rm B}},d_{{\rm B}}$ (see section 4.1). 
Condition (\ref{oi2}%
) gives 
\begin{equation}
\widetilde{b}_{1}=5\widetilde{d}_{1},\qquad \qquad \widetilde{c}_{1}=-\frac{%
39}{30}\widetilde{d}_{1},  \label{anu}
\end{equation}
while 
(\ref{con}) imposes 
\begin{equation}
a_{2}=\frac{25}{6}\widetilde{d}_{1},  \label{ana}
\end{equation}
so that (\ref{auto}) reads 
\begin{equation}
\frac{25}{6(n-6)}\widetilde{d}_{1}\tilde{{\rm G}}_{6}.  \label{fre}
\end{equation}
This is the form of the simple poles. We now show that the finite parts (as
well as the other poles) are related in a similar way. Formula (\ref{ana})
is the analogue of the identification $a=a^{\prime }$ of ref. \cite{athm}.

We compare the radiative correction to the coefficient $-\beta
_{a}$ of ${\rm G}_{6}$ with the prediction (\ref{prediction}). We recall that 
\[
-\beta _{a}={\frac{1}{2\lambda }}{\frac{\partial }{\partial \lambda }}\left(
\lambda ^{2}a_{1}\right) ,
\]
according to (\ref{penu}). The same renormalization-group equations (\ref
{penu}) give 
\begin{equation}
-\beta \frac{\partial a_{1}}{\partial \lambda }=\frac{1}{2\lambda }{\frac{%
\partial }{\partial \lambda }}\left( \lambda ^{2}a_{2}\right) .  \label{aa}
\end{equation}
Here, $a_{2}={\frac{25}{6}}\tilde{d}_{1}$ can be read from (\ref{puppo}).
The additive constant in $a_{1}$ is not fixed by our procedure. It is the
free-field value of the anomaly, which can however be computed directly \cite
{ichinose}. Combining the above results we have 
\begin{equation}
\Theta =\left( {\rm const.}+{\frac{41}{7464960}}{\frac{\lambda ^{6}}{(4\pi
)^{12}}}\right) {\rm {G}}_{6}+{\rm rest},  \label{reft}
\end{equation}
to the fourth-loop order included, as we wished to show.

Recalling the procedure with which $\tilde{d}_{1}$ was calculated, this
result shows the direct relationship between the coefficient of the Euler
density (which contributes to the $\Theta $ four-point function, since it is
cubic in the Riemann tensor) and the $\Theta $ two-point function, in
agreement with the ideas of \cite{athm}. This relationship in non-trivial
and would not be naively expected to hold. Only when analysing the induced
effective action for the conformal factor does one uncover that a very
simple positivity property is actually equivalent to the claimed
relationship to all-orders in perturbation theory \cite{athm}. We
stress that the positivity arguments of \cite{athm} do apply here
and imply that the induced action for the conformal factor is 
negative definite throughout the RG flow.

We now comment on the remaining terms of (\ref{reft}). The terms of the list
(\ref{three}) should be proportional to the beta function, by the
integrability condition, while the terms of the list (\ref{three2}) should
reconstruct ${\rm {\tilde{G}}}_{6},$ as in (\ref{fre}). However, this is not
straightforward, due to the renormalization
mixing betwen $\Theta $ and $\Box [\varphi ^{2}]$. 
Having already obtained the desired result, namely (\ref{reft}), we will not
pursue this matter further. Nevertheless, we conclude this section by
showing that, consistently with what we have just said, there would be no
trouble in the absence of the $\Box [\varphi ^{2}]$-mixing. In this case,
the equations reported in the appendix (as well as our discussion so far)
would simplify enormously: $\eta ,e,\beta _{\eta },\kappa ^{(1,2,3)},\Lambda
^{(1,2,3)},\Xi ^{(2)},g^{(1,2,3,4)},h^{(1,2,4)}\rightarrow 0$. We are left
with $a,b,c,d$ and $f^{(1,2,3,4)}$, related via an oversimplified version of
(\ref{appul}) and (\ref{penu}), namely --- for $n=6$ --- 
\begin{eqnarray}
5\beta _{b} &=&\frac{\beta }{\lambda }f^{(1)},\qquad 2\beta _{b}+15\beta
_{c}+\beta _{d}=\frac{\beta }{\lambda }f^{(2)},  \nonumber \\
10\beta _{d} &=&\frac{\beta }{\lambda }f^{(3)},\qquad 4\beta _{b}=\frac{%
\beta }{\lambda }f^{(4)},\qquad D(b,c,d)=\beta \frac{\partial (b,c,d)}{%
\partial \lambda }=\beta _{b,c,d}.  \label{ceffe}
\end{eqnarray}
We see that $\beta _{b,c,d}$, coefficients of the terms (\ref{three}) in $%
\Theta $, are proportional to the beta function, in agreement with the
integrability condition. The coefficients of the terms (\ref{three2}), 
\[
-(f^{(1)}+10b)\Box R_{\mu \nu }R^{\mu \nu }-\left( f^{(2)}+4b+30c+{2}%
d\right) \Box R^{2}-(f^{(3)}+20d)\Box ^{2}R-(f^{(4)}+8b)\nabla _{\mu
}(R^{\mu \nu }\nabla _{\nu }R),
\]
can be found by repeating the above procedure for $\beta _{a}.$ We have,
using (\ref{ceffe}),
\[
-\frac{1}{2\lambda }\frac{\partial }{\partial \lambda }\left\{ \lambda
^{2}\left[ 10b\Box R_{\mu \nu }R^{\mu \nu }+(4b+30c+{2}d)\Box R^{2}+20d\Box
^{2}R+8b\nabla _{\mu }(R^{\mu \nu }\nabla _{\nu }R)\right] \right\} .
\]
Now, (\ref{penu}) gives also equations that are analogous to (\ref{aa}): 
\[
-\beta \frac{\partial b}{\partial \lambda }=\frac{1}{2\lambda }{\frac{%
\partial }{\partial \lambda }}\left( \lambda ^{2}b_{1}\right) ,\qquad -\beta 
\frac{\partial c}{\partial \lambda }=\frac{1}{2\lambda }{\frac{\partial }{%
\partial \lambda }}\left( \lambda ^{2}c_{1}\right) ,\qquad -\beta \frac{%
\partial d}{\partial \lambda }=\frac{1}{2\lambda }{\frac{\partial }{\partial
\lambda }}\left( \lambda ^{2}d_{1}\right) .
\]
Therefore, the relations (\ref{anu}) and (\ref{ana}) between $a_{2}$ and $%
b_{1},c_{1},d_{1}$ are promoted to similar relations between $a_{1}$ and $%
b,c,d$ (up to the usual additive constants), so that, in turn, (\ref{reft})
is promoted to 
\[
\Theta ={\frac{1}{2\lambda }}{\frac{\partial }{\partial \lambda }}\left(
\lambda ^{2}a_{1}\right) {\rm {\tilde{G}}}_{6}+{\rm rest.}
\]

\section{General dimension}

We now discuss the case of generic $n$. The pondered Euler density has the
form 
\begin{equation}
\tilde{{\rm G}}_{n}={\rm G}_{n}+\nabla _{\alpha
}J_{n}^{\alpha }={\rm G}_{n}+\cdots +p_{n}\Box ^{n/2-1}R,\qquad \qquad
J_{n}^{\alpha }=\cdots +p_{n}\nabla ^{\alpha }\Box ^{{\frac{n}{2}}-2}R.
\label{pondenn}
\end{equation}
The dots stand for a list of trivial total derivative terms, which are at
least quadratic in the curvature tensor. The weights of the various terms in
the list are chosen in such a way that on conformally-flat metrics
\[
\sqrt{g}\tilde{{\rm G}}_{n}=-2(n-1)p_{n}\Box ^{n\over 2}\phi .
\]
Only the coefficient $p_{n}$ in (\ref{pondenn})
is relevant to quantum irreversibility and the definition of $\tilde{{\rm G}}%
_{n}$ makes it easily calculable. 
We plan to devote a separate paper to the general mathematical constrution of $\tilde{{\rm G}}_{n}$. 

The computation of $p_{n}$ proceeds as follows. The Euler characteristic of
an $n$-dimensional sphere $S^{n}$, equal to 2, can be written as 
\[
2={\frac{(-1)^{\frac{n}{2}}}{2^{\frac{3n}{2}}\pi ^{\frac{n}{2}}\left({
n\over 2}\right)!}}%
\int_{S^{n}}\sqrt{g}{\rm G}_{n}\,{\rm d}^{n}x={\frac{(-1)^{\frac{n}{2}}}{2^{%
\frac{3n}{2}}\pi ^{\frac{n}{2}}\left({
n\over 2}\right)!}}\int_{S^{n}}\sqrt{g}\tilde{{\rm G}}_{n}\,%
{\rm d}^{n}x=-{\frac{2(-1)^{\frac{n}{2}}(n-1)}{2^{\frac{3n}{2}}\pi ^{\frac{n%
}{2}}\left({
n\over 2}\right)!}}p_{n}\int_{S^{n}}\Box ^{\frac{n}{2}}\phi \,{\rm d}^{n}x,
\]
where 
\begin{equation}
{\rm G}_{n}=(-1)^{n\over 2}
\varepsilon _{\mu _{1}\nu _{1}\cdots \mu _{\frac{n}{2}}\nu _{%
\frac{n}{2}}}\varepsilon ^{\alpha _{1}\beta _{1}\cdots \alpha _{\frac{n}{2}%
}\beta _{\frac{n}{2}}}\prod_{i=1}^{\frac{n}{2}}R_{\alpha _{i}\beta
_{i}}^{\mu _{i}\nu _{i}}.
\label{definition}
\end{equation}
A sphere has the metric ${\rm d}s^{2}={\frac{{\rm d}x^{2}}{(1+x^{2})^{2}}}$,
so that $\phi =-\ln (1+x^{2})$. A tedious, but straightforward, computation
gives 
\[
\Box ^{\frac{n}{2}}\phi =2^{\frac{3n}{2}-1}(-1)^{\frac{n}{2}}\Gamma \left( 
\frac{n}{2}\right) (n-1)!!\frac{1}{(1+|x|^{2})^{n}}
\]
and 
\[
\int \Box ^{\frac{n}{2}}\phi =2^{n}(-1)^{\frac{n}{2}}\pi ^{\frac{n}{2}%
}\Gamma \left( \frac{n}{2}\right) .
\]
We have therefore the result 
\[
p_{n}=-{\frac{2^{\frac{n}{2}}n}{2(n-1)}},
\]
which agrees with the known values in $n=4$ (\ref{formula}) and $n=6$ (\ref
{jeia}). With the help of a computer we have checked this formula for $n=8$,
where the pondered Euler density reads
\[
\tilde{{\rm G}}_{8}={\rm G}_{8}-\frac{64}{7}\Box ^{3}R+\nabla _{\alpha }J_{8~%
{\rm red}}^{\alpha },
\]
with
\begin{eqnarray*}
J_{8~{\rm red}}^{\alpha } &=&\frac{1984}{147}\left( \nabla ^{\alpha }\Box
R\right) R-\frac{1280}{21}\left( \nabla _{\mu }\Box R\right) R^{\mu \alpha }+%
\frac{1024}{63}\left( \nabla _{\rho }R^{\alpha \mu }\right) R_{\mu \nu
}R^{\nu \rho }+\frac{21376}{441}\left( \nabla ^{\alpha }R_{\mu \nu }\right)
R^{\mu \nu }R \\
&&-\frac{256}{3}\left( \nabla _{\mu }R_{\rho \sigma }\right) R^{\rho \sigma
}R^{\mu \alpha }-\frac{9728}{1029}\left( \nabla ^{\alpha }R\right) R^{2}+%
\frac{74752}{3087}\left( \nabla _{\mu }R\right) RR^{\mu \alpha } \\
&&-\frac{4800}{7}\left( \nabla ^{\alpha }R^{\mu \nu }\right) \nabla _{\mu
}\nabla _{\nu }R-\frac{256}{3}\left( \nabla ^{\mu }R^{\nu \rho }\right)
\nabla ^{\alpha }\nabla _{\rho }R_{\mu \nu }+\frac{64}{7}\left( \nabla _{\mu
}R\right) \Box R^{\mu \alpha } \\
&&+\frac{2752}{3}\left( \nabla ^{\alpha }R_{\mu \nu }\right) \Box R^{\mu \nu
}+\frac{101120}{147}\left( \nabla ^{\nu }R^{\mu \alpha }\right) \nabla _{\mu
}\nabla _{\nu }R-\frac{23552}{21}\left( \nabla ^{\nu }R^{\mu \alpha }\right)
\Box R_{\mu \nu }.
\end{eqnarray*}

On conformally-flat metrics we have 
\[
\sqrt{g}\tilde{{\rm G}}_{n}=2^{\frac{n}{2}}n\,\Box ^{\frac{n}{2}}\phi 
\]
for $n\geq 2$. For $n=2$, ${\rm G}_{2}=\tilde{{\rm G}}_{2}$.

We are ready to write the general formula for the $a$-flow. We normalize $a$
in such a way that the trace anomaly equation reads at criticality 
\[
\Theta =a_{n}\,\tilde{{\rm G}}_{n}+{\rm conformal\,invariants}=2^{\frac{n}{2}%
}n\,a_{n}\,{\rm e}^{-n\phi }\Box ^{\frac{n}{2}}\phi .
\]
The two-point function reads 
\[
\langle \Theta (x)\,\Theta (y)\rangle =-2^{{\frac{n}{2}}}n\,a_{n}\,\Box ^{%
\frac{n}{2}}\delta (x-y)
\]
and the expression for the $a$-flow is:
\begin{equation}
a_{n}^{{\rm UV}}-a_{n}^{{\rm IR}}={\frac{1}{2^{{\frac{3n}{2}}-1}\,n\,\Gamma
(n+1)}}\int {\rm d}^{n}x\,|x|^{n}\,\langle \Theta (x)\,\Theta (0)\rangle ,
\label{generaln}
\end{equation}
using $\Box ^{\frac{n}{2}}|x|^{n}=n\,(2n-2)!!$.

Using \cite{ichinose}, we can read the values of $a_{n}$ for free fields in
arbitrary dimension and
normalize $a_{n}$ so that it equals 1 for a real scalar field, and
reads in general 
\begin{equation}
N_{s}+f_{n}N_{f}+v_{n}N_{v}  \label{norma}
\end{equation}
for free-field theories.

The odd-dimensional formula of quantum irreversibility that we propose is
obtained by putting $n=$odd in (\ref{generaln}). For example,
\[
a_{{\rm UV}}-a_{{\rm IR}}={\frac{1}{144\sqrt{2}}}\int {\rm d}^{3}x\,|x|^{3}\,\langle \Theta (x)\,\Theta (0)\rangle
\]
in three dimensions. The normalization (\ref
{norma}), as well as the free-field values of $a$, can be extended
to odd dimensions in the same way. This recipe {\it defines} the $a$%
-function in odd dimensions. The value of $a$ for an interacting
odd-dimensional conformal field theory connectible
to a free-field theory by an RG flow is defined by formula 
(\ref{generaln}) itself.

This prescription has a predictive content. It can be checked,
in principle, in the case of an RG flow interpolating between two
free-field theories, where $a_{n}^{{\rm UV}}-a_{n}^{{\rm IR}}$
can be worked out in two independent ways: the evaluation of (\ref{generaln}) and the continuation of 
the free-field value of $a$, for example (\ref{norma}).

There is no known odd-dimensional example of an RG flow between two free-field
theories. Its existence is plausible, however, since four-dimensional
situations of this type are known, for example the lower bound of
the conformal window in supersymmetric QCD, which, according to
non-Abelian electric--magnetic duality \cite{seiberg}, is manifestly free in a dual formulation.

\section{Conclusions}

A ``pondered'' notion of Euler density \~{G}, linear in the conformal factor, 
is naturally singled out by the
radiative corrections to the trace anomaly in external gravity and is
relevant to the phenomenon of quantum irreversibility, measured by the
invariant area of the graph of the beta function between the fixed points.
The existence of \~{G} is a support to the ideas of ref. \cite{athm}.
Formulas for arbitrary even dimension have been derived, and detailed
calculations in six and eight dimensions have been performed to check the
predictions. A natural, odd-dimensional formula has been
proposed by dimensional continuation.

Our expression for $\Delta a$ can be checked order by order in
perturbation theory, even when there is no interacting fixed point. Here,
the $\varphi ^{3}$-theory in six dimensions has been considered in detail.
Complications are due to the large number of invariants, as well as the
renormalization mixing between the operators $\varphi ^{3}$ and $\varphi ^{2}
$. In higher dimensions it should be 
possible to test our ideas using
higher-derivative renormalizable field theories as a laboratory. We recall, anyway, that the idea of
quantum irreversibility is meaningful in complete generality, even if the
theory is non-renormalizable, provided that the intrinsic RG\ running of the
coupling constants is isolated (in a way to be uncovered) from the spurious effects of dimensioned
parameters \cite{athm}. The inequality $a_{{\rm UV}}\geq a_{{\rm IR}}$ might
be violated, since higher-derivative theories are not unitary, but formula (%
\ref{generaln}) should hold, at least if suitable positivity restrictions
are imposed on the classical action.

\subsection*{Appendix}

I report in this appendix some useful formulas, starting from the
expressions of the curvature tensors in a conformally flat metric, 
\[
g_{\mu \nu }={\rm e}^{2\phi }\delta _{\mu \nu }~~~~~~~~~~~~~~~~~~~~~\Gamma
_{\mu \nu }^{\rho }=\delta _{\nu }^{\rho }\partial _{\mu }\phi +\delta _{\mu
}^{\rho }\partial _{\nu }\phi -\delta _{\mu \nu }\partial ^{\rho }\phi 
\]
\begin{eqnarray}
R &=&-2(n-1)\,{\rm e}^{-2\phi }\left[ \Box \phi +\frac{n-2}{2}(\partial
_{\mu }\phi )^{2}\right] ,  \nonumber \\
R_{\mu \nu } &=&-(n-2)\partial _{\mu }\partial _{\nu }\phi -\delta _{\mu \nu
}\Box \phi +(n-2)\partial _{\mu }\phi \partial _{\nu }\phi -(n-2)\delta
_{\mu \nu }(\partial _{\alpha }\phi )^{2}.
\end{eqnarray}
\[
\Box _{{\rm cov}}S={\rm e}^{-2\phi }(\Box +(n-2)\partial ^{\alpha }\phi
\partial _{\alpha })S,~~~~~~~~~~~~S={\rm scalar}.
\]
The meaning of the symbol $\Box $ (covariant or flat-space d'Alembertian) is
clear from the context in which it is used. The conformal variations needed in
the paper are 
\[
\delta R=-2(n-1)\Box \delta \phi -2R\delta \phi ,~~~~~~~~~~~~~~~\delta
R_{\mu \nu }=-(n-2)\nabla _{\mu }\nabla _{\nu }\delta \phi -g_{\mu \nu }\Box
\delta \phi ,
\]
\begin{eqnarray}
\delta \int R^{3} &=&(n-6)R^{3}-6(n-1)\Box R^{2}  \nonumber \\
\delta \int RR_{\mu \nu }R^{\mu \nu } &=&(n-6)RR_{\mu \nu }R^{\mu \nu }-{%
\frac{n+2}{2}}\Box R^{2}-2(n-1)\Box (R_{\mu \nu }R^{\mu \nu })-2(n-2)\nabla
_{\mu }(R^{\mu \nu }\nabla _{\nu }R)  \nonumber \\
\delta \int R_{\mu \nu }R^{\nu \rho }R_{\rho }^{\mu } &=&(n-6)R_{\mu \nu
}R^{\nu \rho }R_{\rho }^{\mu }+{\frac{3}{4}}{\frac{n-2}{n-1}}\Box R^{2}-{%
\frac{3}{2}}n\Box (R_{\mu \nu }R^{\mu \nu })-{\frac{3}{2}}{\frac{n(n-2)}{n-1}%
}\nabla _{\mu }(R^{\mu \nu }\nabla _{\nu }R)  \nonumber \\
\delta \int R\Box R &=&(n-6)R\Box R-{\frac{n-2}{2}}\Box R^{2}-4(n-1)\Box
^{2}R.  \nonumber \\
\delta \int {\rm G}_{6} &=&(n-6){\rm G}_{6}.
\end{eqnarray}
We have used (\ref{iden}) systematically. Here $\delta ={\frac{1}{\sqrt{g}}}{%
\frac{\delta }{\delta \phi }}$, $\phi $ denoting the conformal factor of the
metric.

The renormalized operator $[\varphi ^{3}]$ is obtained starting from the
finite insertions of $\frac{\partial S}{\partial \lambda }$ in correlators.
We describe here some relevant aspects of the procedure.

There is an important preliminary observation to make. When taking the $%
\lambda $-derivative of the action $S,$ we can omit any finite term and keep
just the poles. Indeed, finite additional contributions to $[\varphi ^{3}]$
(having a correct power expansion in $\lambda $) do not affect the result
that we need, since they appear multiplied by $n-6$ in $\Theta $.

Moreover, we have to omit finite terms such as $\frac{\partial a}{\partial
\lambda }{\rm G}_{6},$ $\frac{\partial b}{\partial \lambda }RR_{\mu \nu
}R^{\mu \nu }$, etc., but for a different reason. Although they are finite
(and therefore do not change the pole part of the renormalized operator),
the expressions $\frac{\partial a}{\partial \lambda },$ $\frac{\partial b}{%
\partial \lambda },$ etc., do not have an ordinary power expansion in $%
\lambda $ \cite{hathrell2} and cannot appear in the definition of $[\varphi
^{3}]$. The reason is readily explained. From (\ref{penu}) we see that $a$
satisfies the equation 
\begin{equation}
\beta {\frac{\partial a}{\partial \lambda }}+{\frac{n-6}{2\lambda }}{\frac{%
\partial }{\partial \lambda }}\left( \lambda ^{2}a\right) =\beta
_{a}(\lambda ).  \label{hy}
\end{equation}
A power expansion for $\beta _{a}(\lambda )$, assured by our arguments,
excludes a power expansion for $a$. This is not a contradiction in the
derivation, but an important property of the functions appearing in the
induced effective action for the external gravitational background. The
expressions in question do not appear in $\Theta $, nor in the induced
action projected onto conformally flat metrics.

A more direct explanation of this matter is obtained by considering (\ref{hy}%
) for $n=6$: we see that ${\frac{\partial a}{\partial\lambda}}$ contains a
denominator $1/\beta$ and we run into the problems described in the
introduction. Such terms cannot appear in the expression of $[\varphi^3]$.
Once this observation is kept in mind, there is no further obstruction to
correctly identify the renormalized operator $[\varphi^3]$.

We have therefore, using the RG equations of section 4.2, 
\begin{eqnarray}
\frac{\partial \lambda _{{\rm B}}}{\partial \lambda } &=&\frac{1}{2}\frac{n-6%
}{\widehat{\beta }}\lambda _{{\rm B}},\qquad \qquad \qquad \qquad \frac{%
\partial \ln Z_{1}}{\partial \lambda }=\frac{\gamma }{\widehat{\beta }}, 
\nonumber \\
\frac{\partial \eta _{{\rm B}}}{\partial \lambda } &=&-\frac{1}{\widehat{%
\beta }}(\beta _{\eta }+\delta \eta )Z_{2}^{-1},\qquad \frac{\partial a_{%
{\rm B}}}{\partial \lambda }\rightarrow \mu^{n-6} {\frac{\partial L_a}{%
\partial\lambda}}= -\frac{n-6}{\widehat{\beta }}\mu ^{n-6}\left( L\Sb a  \\ 
\endSb +\frac{\beta _{a}}{n-6}\right) ,  \nonumber \\
\frac{\partial (b,c,d)_{{\rm B}}}{\partial \lambda } &\rightarrow &-\mu
^{n-6}\frac{1} {\widehat{\beta }}\left[ (n-6)(L_{b,c,d}+\eta L_{\kappa
^{(1,2,3)}}+\eta ^{2}L_{\Lambda ^{(1,2,3)}}+\eta ^{3}L_{\Sigma
^{(2)}})\right.  \label{a0} \\
&&\qquad\qquad + \beta _{b,c,d}+\beta_\eta L_{\kappa^{(1,2,3)}}+\eta (\beta
_{\kappa ^{(1,2,3)}}+2\beta_\eta L_{\Lambda^{(1,2,3)}}+\delta
L_{\kappa^{(1,2,3)}})  \nonumber \\
&&\qquad\qquad\left.+\eta ^{2}(\beta _{\Lambda ^{(1,2,3)}}+3\beta_\eta
L_{\Sigma^{(2)}}+2\delta L_{\Lambda^{(1,2,3)}}) +\eta ^{3}(\beta _{\Sigma
^{(2)}}+3\delta L_{\Sigma^{(2)}}) \right]  \nonumber
\end{eqnarray}

This analysis fixes only the integrated renormalized operator and so we can
add arbitrary local terms, finite or divergent, which have to be fixed by
requiring that $\Theta $ be finite and by studying the renormalization-group
equation of $[\varphi ^{3}]$. The possible local terms are (\ref{three2})
plus $\Box [\varphi ^{2}]$. We write their coefficients as shown in eq. (\ref
{ho}).

Finally, the renormalized operator ${\frac{1}{3!}}\mu ^{-(n-6)/2}[\varphi
^{3}]$ turns out to be equal to ${\frac{(n-6)}{\hat{\beta}}}$ times 
\begin{eqnarray}
&&\frac{1}{2}{\frac{\lambda _{{\rm B}}}{3!}}\varphi _{{\rm B}}^{3}-{\frac{1}{%
2}}{\frac{\beta _{\eta }+\eta \delta }{n-6}}R[\varphi ^{2}]+{\frac{\gamma }{%
n-6}}[{\rm E}]-{\frac{e+L_{e}}{n-6}}\Box [\varphi ^{2}]-\mu ^{n-6}\left[
\left( L_{a}+{\frac{\beta _{a}}{n-6}}\right) {\rm G}_{6}\right.  \nonumber \\
&&+\left( L_{b,c,d}+\eta L_{\kappa ^{(1,2,3)}}+\eta ^{2}L_{\Lambda
^{(1,2,3)}}+\eta ^{3}L_{\Sigma ^{(2)}}\right. \qquad  \nonumber \\
&&\qquad \qquad \qquad \qquad \qquad \left. +{\frac{\beta _{b,c,d}+\eta
\beta _{\kappa ^{(1,2,3)}}+\eta ^{2}\beta _{\Lambda ^{(1,2,3)}}+\eta
^{3}\beta _{\Sigma ^{(2)}}}{n-6}}\right) (RR_{\mu \nu }R^{\mu \nu
},R^{3},R\Box R)  \nonumber \\
&&+{\frac{1}{n-6}}\left( f^{(1,2,3,4)}+\eta g^{(1,2,3,4)}+\eta
^{2}h^{(1,2,3)}\right.  \label{a1} \\
&&\qquad \qquad \qquad \left. \left. +L_{f^{(1,2,3,4)}}+\eta
L_{g^{(1,2,3,4)}}+\eta ^{2}L_{h^{(1,2,3)}}\right) (\Box (R_{\mu \nu }R^{\mu
\nu }),\Box R^{2},\Box ^{2}R,\nabla _{\mu }(R^{\mu \nu }\nabla _{\nu
}R))\right]  \nonumber
\end{eqnarray}
One can check that the operator $\frac{1}{3!}[\varphi ^{3}]\mu ^{-(n-6)/2}$
defined by this formula has a correct power expansion in $\lambda $.

Finiteness of $\Theta $, formula (\ref{tttt}), fixes the pole series of the
total-derivative terms in $[\varphi ^{3}]$. We obtain
\begin{eqnarray}
L_{e} &=&-(n-1)L_{\eta },  \nonumber \\
L_{f^{(1)}} &=&-2(n-1)(L_{b}-L_{\eta }L_{\kappa ^{(1)}}),\qquad
L_{g^{(1)}}=4(n-1)L_{\eta }L_{\Lambda ^{(1)}},\qquad
L_{h^{(1)}}=2(n-1)L_{\Lambda ^{(1)}},  \nonumber \\
L_{f^{(2)}} &=&-\frac{1}{2}(n+2)L_{b}-6(n-1)L_{c}-{\frac{n-2}{2}}%
L_{d}+2(n-1)L_{\eta }L_{\kappa ^{(2)}},  \nonumber \\
L_{g^{(2)}} &=&-\frac{1}{2}(n+2)L_{\kappa ^{(1)}}-4(n-1)L_{\kappa ^{(2)}}-{%
\frac{n-2}{2}}L_{\kappa ^{(3)}}+4(n-1)L_{\eta }L_{\Lambda ^{(2)}},  \nonumber
\\
L_{h^{(2)}} &=&-\frac{1}{2}(n+2)L_{\Lambda ^{(1)}}-2(n-1)L_{\Lambda ^{(2)}}-{%
\frac{n-2}{2}}L_{\Lambda ^{(3)}}+6(n-1)L_{\eta }L_{\Sigma ^{(2)}},  \nonumber
\\
L_{f^{(3)}} &=&-2(n-1)(2L_{d}-L_{\eta }L_{\kappa ^{(3)}}),\qquad
L_{g^{(3)}}=-2(n-1)(L_{\kappa ^{(3)}}-2L_{\eta }L_{\Lambda ^{(3)}}), \\
L_{f^{(4)}} &=&-2(n-2)L_{b},\qquad \qquad L_{g^{(4)}}=-2(n-2)L_{\kappa
^{(1)}},\qquad \qquad L_{h^{(4)}}=-2(n-2)L_{\Lambda ^{(1)}},  \nonumber
\label{a2}
\end{eqnarray}
Next, we have to relate the finite functions $f,g,h$ to the other finite
functions. This can be done by studying the renormalization-group properties
of the operator $[\varphi ^{3}]$. We observe that $\hat{\beta}^{-1}D\hat{%
\beta}$ produces finite quantities when acting on finite quantities. We
therefore consider 
\[
\hat{\beta}^{-1}D\left( \hat{\beta}{\frac{1}{3!}}\mu ^{-(n-6)/2}[\varphi
^{3}]\right) =({\rm finite}),
\]
re-express $[\varphi ^{3}]$ in terms of $\Theta ,$ and use $D\Theta =0$.
Moreover we use 
\[
D[\varphi ^{2}]=-\delta [\varphi ^{2}]-2\mu ^{n-6}(\beta _{\kappa
^{(1,2,3)}}+2\eta \beta _{\Lambda ^{(1,2,3)}}+3\eta ^{2}\beta _{\Sigma
^{(2)}})(R_{\mu \nu }R^{\mu \nu },R^{2},\Box R),
\]
which can be found by direct inspection of (\ref{varf}). The result is 
\begin{eqnarray}
(n-1)\beta _{\eta }=\delta e,\qquad 2(n-1)\beta _{b}-2e\beta _{\kappa
^{(1)}}+\beta _{\eta }g^{(1)} &=&2\frac{\beta }{\lambda }f^{(1)},  \nonumber
\label{appul} \\
-4e\beta _{\Lambda ^{(1)}}+2\beta _{\eta }h^{(1)}=\left( 2\frac{\beta }{%
\lambda }-\delta \right) g^{(1)},\qquad -(n-1)\beta _{\Lambda ^{(1)}}
&=&\left( \frac{\beta }{\lambda }-\delta \right) h^{(1)},  \nonumber \\
\frac{n+2}{2}\beta _{b}+6(n-1)\beta _{c}+{\frac{n-2}{2}}\beta _{d}-2e\beta
_{\kappa ^{(2)}}+\beta _{\eta }g^{(2)} &=&2\frac{\beta }{\lambda }f^{(2)}, 
\nonumber \\
\frac{n+2}{2}\beta _{\kappa ^{(1)}}+4(n-1)\beta _{\kappa ^{(2)}}+{\frac{n-2}{%
2}}\beta _{\kappa ^{(3)}}-4e\beta _{\Lambda ^{(2)}}+2\beta _{\eta }h^{(3)}
&=&\left( 2\frac{\beta }{\lambda }-\delta \right) g^{(2)},  \nonumber \\
\frac{n+2}{2}\beta _{\Lambda ^{(1)}}+2(n-1)\beta _{\Lambda ^{(2)}}+{\frac{n-2%
}{2}}\beta _{\Lambda ^{(3)}}-6e\beta _{\Sigma ^{(2)}}+\delta \beta _{\Lambda
^{(3)}} &=&2\left( \frac{\beta }{\lambda }-\delta \right) h^{(2)},  \nonumber
\\
4(n-1)\beta _{d}-2e\beta _{\kappa ^{(3)}}+\beta _{\eta }g^{(3)}=2\frac{\beta 
}{\lambda }f^{(3)},\quad 2(n-1)\beta _{\kappa ^{(3)}}-4e\beta _{\Lambda
^{(3)}} &=&\left( 2\frac{\beta }{\lambda }-\delta \right) g^{(3)},  \nonumber
\\
2(n-2)\beta _{b}+\beta _{\eta }g^{(4)}=2\frac{\beta }{\lambda }%
f^{(4)},\qquad 2(n-2)\beta _{\kappa ^{(1)}}+2\beta _{\eta }h^{(4)} &=&\left(
2\frac{\beta }{\lambda }-\delta \right) g^{(4)},  \nonumber \\
(n-2)\beta _{\Lambda ^{(1)}} &=&\left( \frac{\beta }{\lambda }-\delta
\right) h^{(4)}.  \nonumber
\end{eqnarray}

\end{document}